\documentclass[preprint,3p,times,twocolumn,12pt]{elsarticle}
\usepackage{lineno}
\usepackage[utf8]{inputenc}
\usepackage[T1]{fontenc}
\usepackage[english]{babel}
\usepackage{fancyhdr}
\usepackage{calc}
\usepackage{graphicx}
\usepackage{wrapfig}
\usepackage{geometry}
\usepackage[debug]{hyperref}
\usepackage{caption}
\usepackage{amsmath}
\usepackage{amsfonts}
\usepackage{multirow}
\usepackage{amssymb}  
\usepackage{natbib}  
\usepackage[export]{adjustbox}   
\usepackage[utf8]{inputenc}
\usepackage[english]{babel}
\usepackage[dvipsnames]{xcolor}
\usepackage[gen]{eurosym}

\hypersetup{backref=true, 
pagebackref=true,
hyperindex=true, 
colorlinks=true, 
breaklinks=true, 
urlcolor= cyan, 
linkcolor= purple, 
bookmarks=true, 
bookmarksopen=true}
\usepackage{multicol}
\usepackage{array}
\usepackage{soul}
\usepackage{ulem}
\journal{Ecological Economics}

\usepackage{bm}
\newcommand{\bigdot}[1]{\overset{\bm .}{#1}}

\definecolor{cyan}{rgb}{0.2,0.6,1}
\definecolor{red}{rgb}{1,0,0}
\definecolor{blue}{rgb}{0,0,1}
\definecolor{green}{rgb}{0,0.5,0}
\definecolor{black}{rgb}{0,0,0}

\renewcommand{\@}{\textcolor{cyan}}
\renewcommand{\b}{\textcolor{black}}
\newcommand{\g}{\textcolor{black}}

\usepackage{setspace}

\biboptions{authoryear}

\begin{document}

\begin{frontmatter}
\title{Macroeconomic Dynamics in a finite world: the Thermodynamic Potential Approach}

\date{today}

\author[label3]{Éric Herbert\corref{cor1}}
\cortext[cor1]{corresponding author}
\ead{eric.herbert@u-paris.fr}
\author[label1,label2,label5]{Gael Giraud}
\author[label2,label3]{Aurélie Louis-Napoléon}
\author[label3]{Christophe Goupil}

\address[label3]{Laboratoire Interdisciplinaire des Energies de Demain, LIED UMR 8236, Universit\'e Sans Nom, Paris, France}
\address[label1]{Environmental Justice Program, Georgetown University, Washington DC}
\address[label2]{Chaire Énergie et Prospérité, Paris, France}
\address[label5]{Centre d'Economie de la Sorbonne, CNRS, Paris, France}

\begin{abstract}
This paper presents a conceptual model describing the medium and long-term co-evolution of natural and socio-economic subsystems of Earth. An economy is viewed as an out-of-equilibrium dissipative structure that can only be maintained with a flow of energy and matter. The distinctive approach emphasized here consists in capturing the economic impact of  natural ecosystems being depleted and destroyed by human activities via a pinch of thermodynamic potentials. 
This viewpoint allows: $(i)$ the full-blown integration of a limited quantity of primary resources into a non-linear macrodynamics that is stock-flow consistent both in terms of matter-energy as well as economic transactions; $(ii)$ the inclusion of natural and forced recycling; $(iii)$ the inclusion of a friction term which reflects the impossibility of producing goods and services in high metabolising intensity without exuding energy and matter wastes; $(iv)$ the computation of the anthropically produced entropy as a function of intensity and friction. Analysis and numerical computations confirm the role played by intensity and friction as key factors for sustainability. Our approach is flexible enough to allow for various economic models to be embedded into our thermodynamic framework.
\end{abstract}

\begin{keyword}
Ecological macroeconomics \sep Laws of Thermodynamics \sep Recycling \sep Waste \sep Finite resources
\end{keyword}
\end{frontmatter}

\setlength\abovecaptionskip{0.3ex}
\setlength{\floatsep}{12pt plus 2pt minus 10pt}


\section{Introduction}
\label{Introduction}

Most economic models consider Nature as an easy-to-overcome constraint. Unlimited supply of energy  and matter (E\&M in the following) and full
(or almost full) substitutability of the production factors  --- namely labour $L$, capital $K$ and energy $E$ --- are often
 assumed, see \cite{ofce, stiglitz1974growth}. This modelling choice is not appropriate in the environmental
context of a finite world. The consequences of direct and incidental environmental damages \cite{IPCC} of natural ecosystems \cite{MEA}
and the emergence of resource scarcity (like e.g., fossil fuels, metals or rare earth elements) entail to seriously take into account

Refering back to \cite{Soddy}, thermodynamics comes up naturally when dealing with E\&M:
\textit{``The laws [of Thermodynamics] that express the relation
  between matter and energy, govern the rise and fall of political
  systems, the freedom or bondage of societies, 
the movements of commerce and industries, the origin of wealth and
poverty, and the general physical welfare of a people.''}

According to \cite{Sollner}, integrating thermodynamics into economy
can be attempted at various levels: $(i)$ by introducing an energy theory
of value, see \cite{Odumbook};  
$(ii)$ through analogies and metaphors like for instance in \cite{Samuelson,Ayres1994,Saslow,Sousa}; 
$(iii)$ by adding thermodynamical constraints to a neoclassical
environment like, \textit{e.g.}, in \cite{RUTH,Georgescu,Boulding,Mayumi}.

Actually, when applied, Thermodynamics is integrated in a variety of ways in models, but often not in a satisfactory manner (when not used in an improper way). In these approaches, and sometimes despite some vibrant pleas claiming the contrary, the physical world is only marginally taken into account, and almost never considered as such with its own laws, in particular, those of thermodynamics \cite{Kummel,Berg,Dafermos}. A particular mention can be made of the case of ecological modelling, which is sometimes based on thermodynamics but adds principles such as the so-called maximum power principle (MPP) \cite{Odumbook} \cite{Hall} which is not \textit{per se} a thermodynamic principle  \cite{goupil2020}. 

Following \cite{Sousa}, we affirm that economics as a science does not have the complete structure of thermodynamics, and has no reason to have it. On the other hand, the physical world is totally subject to the laws of thermodynamics. The physical world and the economic world being of course connected, it follows that some elements of thermodynamics may seem to belong to the economy in their own right, but this is inaccurate. This is why the model we present makes a very clear distinction between what belongs to the physical world, which subscribes to the laws of physics, and what belongs to the economic world, which subscribes to the laws of economics.

For this reason, our macro model defines separately a physical world and a distinct economic world. We will see in this article how these two areas communicate.

Thermodynamics differs from other physical sciences in that it deals with both quantities (courtesy of the first law) and quality (thanks to the second law). Quality measures  the ability of a given quantity to exist in a dispersed (bad quality, high entropy) or concentrated (good quality, low entropy) form. The arrow of time implies --- and is actually defined by the fact--- that, in order to perform any physical work, one needs to extract energy and matter from concentrated reservoirs and exude wastes in dispersed form.  For example, in thermal machines to which energy is supplied, a fraction of it may be concentrated as \textit{work} on the condition that the other fraction is dispersed as \textit{heat}. Following \cite{Roegen} and his famous assertion "matter maters too", the pair dispersion-concentration applies equally well to matter without the need to invoke an additional thermodynamic law \cite{Georgescu, Missemer}. Any physical process, including economic activities, involves matter and energy, whose dispersion across time is unavoidable. Manufacturing of a structured object therefore requires a supply of energy of high quality to reduce and reverse the dispersion mechanism.  Similarly, the persistence of a physical structure is only possible if energy of sufficiently high quality is provided to avoid its dispersion. This makes it possible, in particular, to define the energy cost of recycling, that is, of re-concentrating a raw matter that has been dispersed by the production process, \cite{szargut_chemical_1989}. The result is that the concentration of a chemical element respectively in its most concentrated version (ore grade in mines...) or in its most dispersed version (earth's crust, oceans, atmosphere), defines a floor and a ceiling of chemical potentials which may be useful, but not sufficient, to envisage an exergetic approach. This point will be further developed below.

Therefore, any sensible modelling of an economic dynamics must not only make explicit the dependence of economic processes upon energy and matter, but also include a frictional or viscosity term, responsible for neither energy nor matter being entirely used up in an economic activity, but a fraction of them being rejected as waste. Thermodynamics dictates that the influence of the frictional of viscous term depends on the operating intensity of the process, just as it does for any physical system. In other words, to produce cleanly is to produce slowly, the reverse assertion being false in general. 
 
The specific features of our model are:
\begin{itemize}
    \item Resource exploitation and resource availability constraints are explicitly included in the modelling, hence linking economy with the physical world.
    \item Recycling flows are introduced, allowing for a partial re-allocation of dispersed resources.
    \item The distinction between quantities and qualities is ensured by the definition of intensive potentials, and finally, exergy evaluation. 
    \item The choice of a formalism close to out-of-equilibrium  thermodynamics and the introduction of the temporal dimension add rhythmic patterns to flows, entailing not only quantitative, but also qualitative changes in the description of flows. 
    \item A modelling structure based on elementary unit, each simulating a single resource, called physical sheet in the following. Every physical sheet  inner scheme shares features of a conversion engine. The physical world is made up of different core-linked sheets, which guarantees a limited growth in modelling complexity proportional to the number of sheets.
    \item The economic world communicates with the physical one via a \textit{demand function} that replaces the traditional \textit{production function} and is only equivalent to it in the case of unlimited resources. In addition, the production for each sheet, is characterized by the friction term that defines the fraction of waste produced at a given metabolising intensity. As a proxy for the inverse of capital, this friction is subject to constant increase, due to erosion, or decrease in the case of investment and technological progress. The lower the friction term, the easier it is to collect and transform resources. This fluidisation of the production process can then naturally lead to the accelerated depletion of resources insofar as the operating intensity can then increase significantly. Conversely, high friction necessarily imposes a low metabolising intensity, and therefore a low production intensity.
\end{itemize}

According to \cite{Bestiaire}, the current model is one of the models that integrate the economy into the ecosystem, but also those that describe the interactions between the economy and the physical limits of the environment. We propose a framework in which E\&M play the central role, through the use of categories derived from thermodynamics. The quantity of resources is in fact considered limited, and recycling flows are made explicit in the model. The total quantity of a conserved quantity is fixed and the explicit conservation of E\&M in the model requires that this quantity is either in the primary resource state, in the waste reservoir, or in end-use products. For each resource, the sheet structure is formally identical and one sheet corresponds to a specific resource. The use of several sheets makes it easy to account for several resources. Thus, even if the elementary structure is simple, making the model more complex by increasing the number of elementary sheets is not a problem. The formalism of the elementary sheet is that of a conversion machine, which embodies some of the categories used in finite time thermodynamics. Intensity, i.e., the speed at which the economy operates, and friction, characterize the state of the production system (reflecting the fact that an economy with a poorly functioning production framework involves a lot of "friction" in its operations). 

For the sake of concreteness, we have chosen to associate a single sheet to a stock-flow coherent (\cite {LavGod} (SFC)) setting and a Goodwin-Keen type of dynamics (\cite {Grasselli}). The economic model used is heuristic: we want to show that our physical sheets can be connected to an economic model in such a way that both spheres react back on each other. For simplicity, the SFC model is composed only of two sectors: households and businesses. At any time, the existing stock of installed capital, when used at full capacity, dictates a certain level of metabolising intensity (provided there are enough labour forces) which, in turn, requires natural resources to be extracted and depleted at a corresponding speed. The availability of resources (captured through the high entropy potential) and the carrying capacity of Earth as a sink for anthropic wastes (captured though the low entropy potential) enable the metabolising intensity to materialize into an  aggregate ``useful work'' ---the output of the whole economic process--- associated with wastes. Part of this work serves as investment in new capital. Meanwhile, installed capital decays while another part of aggregate useful work is just consumed by households: both processes fuel another flow of wastes which feeds the low-entropy potential. The modified installed capital induces in turn a change in the metabolising intensity, etc. 
 
We show that the resulting model is not only SFC in the traditional, macro-economic sense, but also in the resource-flow sense : it conforms to the conservation of matter and energy. On the other hand, it also fulfills a basic ``correspondence principle'' : when natural resources are infinite, our framework boils down to a standard macro-economic model {\it à la} \cite{Grasselli}. Of course, when resources are finite, the interplay between the physical sphere and the anthropic one sharply modifies the dynamics. In particular, an economy entirely based on non-renewable energies and matter ends up in a collapse.

In the remainder of this article, in section\,\ref{ECE} we first recall some basics of thermodynamics. In section\,\ref{conversion},  we describe our physical sheet, based on Thermodynamics-based categories and we apply it to various examples in section\,\ref{casestudy}. Finally, in section\,\ref{SFC}, we connect  a SFC-Goodwin framework with our physical sheet and show the effects of limited resources, friction and intensity.

\section {Thermodynamics and the energy conversion engine}
\label{ECE}
This section shall specify some thermodynamic concepts, used in the
proposed modelling. 
\subsection{First and Second Laws}
The first law of thermodynamics states that energy is conserved, which means that it can neither be created nor destroyed. This first law is also applicable to matter in the sense that in a closed system, matter is conserved. This naturally applies to the earth \cite{Sollner}: the respective total amounts of carbon, copper, iron, rare earth elements... cannot change significantly over time.\footnote{On the other hand, the earth is not an isolated system since it receives a very large flow of energy from the sun.} The second law states that, if the total amount of matter
cannot change, a change in quality is however possible and {\text thermodynamics essentially describes the transformations from one quality to another}, \cite{Mayumi}. During so-called spontaneous transformations, energy changes quality from a concentrated form, called work, to a diluted form, called heat. The same law affects matter, which, over time, spontaneously degrades, also passing from a concentrated form to a diluted one. Such an observation does not justify invoking a new principle of thermodynamics as has sometimes been advocated. Indeed, changes in the quality of matter are already included in the second law.  
The production processes of consumption goods make use of matter that can be described as concentrated, compared to what happens to this matter when it is then diluted in the waste we generate both during the production and consumption processes.

Recycling matter is therefore in itself a process of concentrating the matter to make it usable again as a resource. For each element and matter resource, we define a potential characterizing its more or less concentrated state. In the case of mineral resources, the concentrated form corresponds to the matter concentrated in the mine, while the diluted form refers to the matter dispersed in the earth's crust, oceans or atmosphere.  As a result, the energy cost to reconcentrate a diluted species is perfectly calculable and used in ecological economics \cite{valero_inventory_2010,szargut_chemical_1989}.

\subsection{Intensive, extensive property}
Another thermodynamic concept is the distinction between extensive and intensive properties. An extensive property depends on the quantity of matter in the system:
when dividing a system in two equal parts, each half is endowed with half the amount of the considered property (like mass or volume for instance). 
An intensive property is independent of the amount of matter \cite{Atkins}. Examples of intensive properties are temperature, pressure, chemical and electrochemical potentials. As they serve to define the thermodynamic potentials, intenstive properties are essential for the mathematical formulation of the second principle and its application in terms of exergy.

\subsection{Thermodynamics conversion and boundary conditions}
\label{Conv}
Living systems, such as organisms and ecosystems, are physical systems, working under non equilibrium conditions. They constitute a specific part of the large family of the thermodynamic systems, from which the Carnot engine is the most famous. In the spirit of \cite{Schrodinger1944, Morowitz1968, Morowitz1978, Schneider-Kay1994} and other authors, we shall consider the physical world as a thermodynamic conversion system, which, like a living organism, metabolises E\&M.

In the economic literature, \cite{Boulding} introduced the distinction  between ``cowboy'' and ``spaceman'' viewpoints. 
In the former, the earth possesses unlimited reservoirs of E\& M: production and consumption are never are considered as physical constraint, usually because technology, monetary capital or labour are postulated to be substitutable to environmental resources (see \textit{e.g.}. 
For a ``spaceman'' economy, reservoirs are limited, as well as waste capacity and no human activity can overcome these limitations. Our approach is in the ``spaceman'' spirit. We insist that the biosphere does not exchange matter with the rest of the universe, but only energy. Production and consumption flows of matter are hence limited in quantities. However, qualities can be
improved if there is a sufficient high quality energy supply. The same idea can be found  in \cite{Mayumi} ``... energy which is sourced from a stock, like oil, can only be used once.  Energy which is sourced from a flow, like solar or geothermal energy, cannot be used at a rate exceeding the source flowrate''. 
Before the late 18th century, economies were organic economies \cite{Wrigley}: the societies were depending on the annual amount of photosynthesis conversion in plants.  Economies were hence ``flow economies'', limited by the solar energy flow. Since the Industrial Revolution, economies are mostly based on fossil fuels, and depend on their available stock: they are ``stock economies''. As a consequence, the nature of the boundary conditions at the edges of the human sphere ---flow \textit{vs} stock conditions--- leads to very different outcomes for an economy. Under  flow conditions, divergence towards an economic metabolism operating at an ever-increasing intensity is impossible because it is limited by the flow of energy supply. Under stock conditions, any improvement in the conversion process is made possible by the energy stock (until it is depleted). As a result, under stock conditions, the trend towards continuous growth in conversion intensities becomes possible. This is in short our current situation, and makes it necessary to consider how the finite size of resources forces and enables us to bifurcate from this trajectory. 

Thermodynamics Laws have two implications in economy \cite{Mayumi,Baum,Sollner}. Firstly, if E\&M are both coming from a stock (e.g. the respective mass of matters in the Earth, the energy contained in fossils fuels), with no recycling, they both can be exhausted. The energy coming from the Sun does not stem from a (human-sized) stock, but is limited by the total solar energy flux the Earth receives. Consequently, the ability of the human sphere to divert energy from biomass and oceans is limited.\footnote{This limitation can be captured, e.g., courtesy of an aggregate parameter such as {\sc hanpp} (Human Appropriation of Net Primary Productivity, in Gt of carbon per annum)\cite{Vitousek,Wright,Haberl}, which is currently close to $\sim 40\%$ and is obviously bounded from above.}
Second, since neither matter, nor energy can be destroyed, the resulting waste has to appear somewhere. Hence, thermodynamics constrains economics: $(i)$ by inputs availability and $(ii)$ by outputs sink capacity. This means not only that, say, $(i)$ the (non-conventional) peak oil or the extraction peak of cupper (\cite{Giraud-Vidal}) will have a huge impact on the capacity of the world economy to keep increasing its {\sc gdp} but also $(ii)$ climate change and biodiversity erosion, which can both be viewed as arising from the limited capacity of environmental sinks to absorb our wastes.

\subsubsection{Efficiency and production}

Consider now a conversion engine able to convert the resources E\&M into goods and services, $G$. We shall consider that the potential of the well (of resources) is high while the potential of the sink (where wastes accumulate) is low. In a thermodynamical system, the potentials thus introduced are intensive variables, corresponding \textit{e.g.} to specific Gibbs chemical potentials. Consequently we define respectively $\mu_H$ and $\mu_L $ for the potential of the well and the sink.

\begin{figure}
	\begin{center}
		\includegraphics[width=0.5\textwidth, trim = 0 0 0 0.7cm]{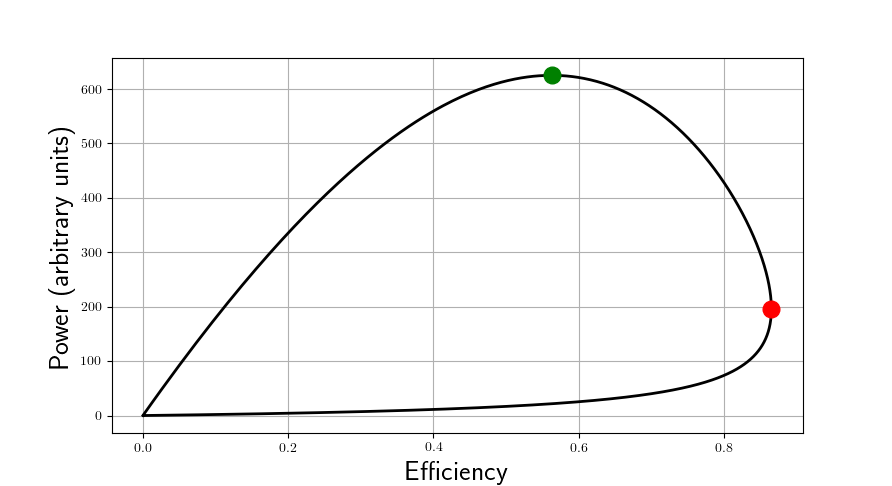} 
		\captionof{figure}{Power in function of efficiency of an energy conversion  engine at constant heat sources temperature, \textit{i.e.} $\Delta \mu$ is constant. Increasing the working intensity from 0, maximum efficiency is reached first (red circle) and then the maximum power (green circle). }
		\label{lobes}
	\end{center}
\end{figure}

High potentials account for the capacity of raw inputs to yield end use goods and services, while the same resources lower the potential difference when contributing to the waste flux.
We shall therefore  relate potentials respectively to the available quantity of resources and to the secondary residual (the ``waste'') stock. For a conversion engine,  input and output are usually thought of as energies, and efficiency can be defined as the output work to input energy ratio. A representation of such a heat conversion process is given in Fig.\,\ref{lobes}. We can show, using the finite time thermodynamics framework \cite{Apertet2}, that power and  efficiency of production are functions of intensity, with heat power as typical resource and sink, and mechanical power as output.  
It is therefore possible to plot one against the other as shown in Fig.\,\ref{lobes}. At low intensity, efficiency and power increase proportionally. Then the curve flexes and the maximum efficiency point is reached. If intensity increases further, then the maximum production point is reached, at the cost of a significant decrease in efficiency. Beyond this point, efficiency and production power collapse. The conversion machine is then approaching exhaustion condition, in the metabolic sense of the term. Each sheet of the physical world composing our model is supposed to work between the points of maximum efficiency and maximum power. However, the economy-based control proposed in this model makes it possible to force the production intensity at will. 

\subsection{Useful work and dissipation }
A substantial literature has shown that living ecosystems can indeed be viewed as out-of-equilibrium systems and, more precisely, as dissipative structures that can only be maintained with a flow of E\&M, \cite{prigogine1997}. At the scale of the complete biosphere, this system works as a thermodynamic thermal engine  which follows cycles of transformation using the sun as a primary hot source and the night sky as an ultimate cold source (see, \textit{e.g.}, \cite{virgo2011} and the references therein). All the living systems of the biosphere are located in the trophic chain, from photosynthetic vegetal receiving their energy from the solar flow, to herbivores that feed from solar energy stored into the  vegetal matter as chemical energy, to carnivores that benefit from a very concentrated chemical energy stored in the meat. From there, one can provide some accounting of matter relying on energy units.

The union of two dissipative structures being still a dissipative structure, a living human society can be viewed as well as an out-of-equilibrum dissipative structure: it receives E\&M in the form of biomass, raw matter, fossil energy, geothermal energy wind and sun light and converts it into work and wastes. Even more, it {\it must} constantly produce some work in order to maintain, or increase, the complexity of its physical structure. Following \cite{ayres2009} and \cite{ayres_economic_2010}, one metric for the type of work performed by human societies is provided by {\it useful work}. The latter focuses on primary energy converters, grouped into the following four categories:
\begin{enumerate}
\item Muscle Work 
\item Electricity, which is electrochemical energy
\item Mechanical Drive
\item Heat (low, mid, and high temperatures)
\end{enumerate}
Useful work is therefore a generalization of mechanical work not to be confused with {\sc gdp} (or national income) measures. It includes both manufactured goods and services but also underground economic and non-monetary transactions. Loosely speaking, one could argue that 
 useful work captures the physical {counterpart of what the economic concept of {\sc gdp} stands for. It must be stressed, however, that even ``real {\sc gdp}'' is a monetary concept which depends upon a given price system while useful work is a purely physical quantity. In the sequel, $G$ stands for the transformed fraction of a given flow of E\&M that contributes to useful work. The flow, $G$, is directly related to the incident and outgoing flows named  $F_{HP}$ and $F_{LP}$ and their associated high and low potentials, $\mu_H$ and $\mu_{L}$.

The incoming and outgoing flows, $F_{HP}$ and $F_{LP}$, and their associated potentials, $\mu_H$ and $\mu_L$, define the incoming entropy flux $\bigdot{S}_H=\frac{F_{HP}}{\mu_H}$, as well as the outgoing  one $\bigdot{S}_L=\frac{F_{LP}}{\mu_L}$. Under ideal reversible conditions, also called the Carnot conditions, we get $\bigdot{S}_H=\bigdot{S}_L$, which means that efficiency reaches its maximum $\eta = \frac{F_{HP}-F_{LP}}{F_{HP}}=1-\frac{\mu_L}{\mu_H} $, obviously lower than unity. Notice that the maximal efficiency of a conversion process never coincides with its maximal instantaneous production. For the most efficient systems --- with efficiency being measured by the ratio between production and resources--- the intensities required for maximal efficiency and maximal production are very different. Paradoxically, only poorly efficient systems exhibit some proximity between these two maxima. In other words, it is very difficult for a system to run at its maximal efficiency and then switch to  its maximal production. This observation is well known and theoretically described as the trade-off between adaptation and adaptability as reported by \cite{goupil2020} (see Fig.\,\ref{lobes})}.

\section{Conversion engine in the economic context}
\label{conversion}

\begin{figure}
	\begin{center}
		\includegraphics[width=0.5\textwidth]{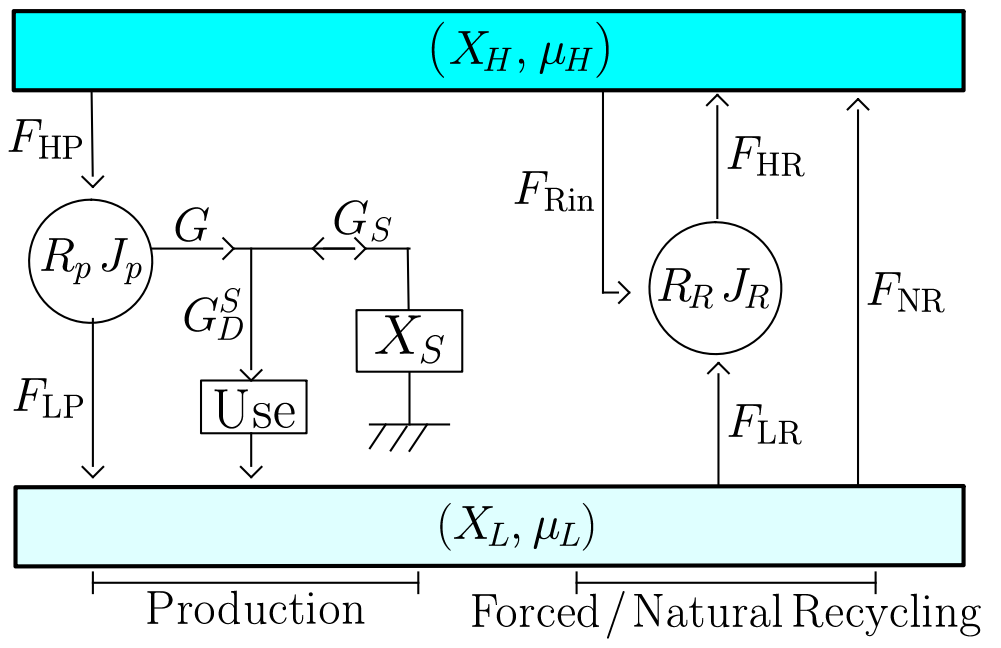}
	\end{center}
	\captionof{figure}{Global scheme of a sheet of resource. Left part is the production area, middle part is forced recycling, and right part is natural recycling.}
	\label{schemacomplet}
\end{figure}

The global scheme features of an economy is depicted in Fig.\,\ref{schemacomplet}. 
The stock, $X_H$, of primary (raw) resources is represented at the top of the diagram. Its availability as an input is accounted for by a higher potential. 
At the bottom, the used resource, $X_L$, denotes the secondary residuals or ``waste''. Its lower potential accounts for its output sink capacity. 
Primary resources (E\&M) as well as flows and residuals throughout the economy are measured in physical units and not in monetary terms. 

\subsection{Production sphere}
The subsheet on the left side of Fig.\,\ref{schemacomplet} provides a schematic for the production sphere. A flux,  $F_{HP}$, of raw resources \b{(regenerating or non-renewable E\&M such as coal, oil, gas, nuclear, biomass, soil, minerals ...)}, which stock is denoted $X_H$, enters the economic metabolism, represented by the ellipse, with a given
quality and availability captured  by the (high) potential, $\mu_H$. 

The economy operates as a conversion engine, producing a flow of useful work, $G$, and rejecting a flow, $F_{LP}$, of  residuals (emissions or wastes) into the sink (atmosphere, ocean, land, forest...), represented by a stock, $X_L$, with a low potential $\mu_L$. The flow $G$ refers to the E\&M transformed by the manufacture of final goods and services. A flux of degradation, $R_PJ_P^2$, accounts for the internal friction of the conversion engine to be specified below. 
 
{\color{black}{The flux, $G$, of final goods can either directly satisfy aggregate demand, or fill a stock, $X_S$ , of inventories (named buffer stock) with flux, $G_S$ , if the production flux is larger than the current demand, $G_D$. Note that $G$ refers to the E\&M transformed for the manufacture of final goods and services. While a fraction of $G$ can be piled in a stock of inventories, there is, as we shall see, no stock of final goods \& services. Of course, aggregate demand, $G_D$, draws both on the buffer stock, $X_S$, and on the current production $G$. In both cases, the incoming E\&M are converted before being rejected to $X_L$.}}

The intensity, $J$, is the global intensity of economic activity but, for each sheet, a specific intensity $J_P=n_P \,  J$, is also defined proportionally to the global intensity. $J_P$ defines the rate at which material resources are converted.\footnote{For $n_P > 1$, the subsector under scrutiny works at a larger intensity than the global intensity of the economy while for $n_P < 1$, the subsector is working less intensely.}
The incoming flux $F_{HP}$ entering the conversion sphere is the product of $J_P$ by the high potential $\mu_H$. Each sheet is assigned a characteristic response time, $\tau>0$, that acts as a low-pass filter against too fast changes in economic intensity. It follows that the operating intensity of a sheet is governed by a differential equation whose solution yields $J_P$, the intensity of the sheet in response to the $J_{P}^D$ intensity requested by the overall economic sphere. Indeed, viewed as part of a physical metabolism, no production sector can react instantly to a change of intensity. The model is based on the work developed by Lars Onsager \cite{onsager1931} on the locally linear thermodynamic response.

Note, in particular, the analogical formulation with that of thermal machines in which the incident energy flows are proportional to the entropy potentials and flows. In the present case, the flow $F_{HP}$ is proportional to $\mu_H$ and to the intensity, $J_P$, which is a proxy of the entropy transport.\footnote{The interested reader will find the basis for this approach in \cite{apertet2013}.}
The equations governing the functioning of a sheet are therefore,
\begin{eqnarray}
X_T     &=& {\color{black}{X_H+X_L+ X_S }} \label{mass} \\ 
J_P     &=& n_P  \, J \label{Intensityprod}\\
J_{P}^D &=& \tau  \, \bigdot{J}_P+ J_P \label{IntensityTau}\\
F_{HP}  &=&\mu_H  \, J_{P} \label{eq:FHP}\\ 
F_{LP}  &=&\mu_L \,  J_{P} + {R_{P}} \,  J_{P}^2 \label{eq:FLP}\\
G       &=& F_{HP} - F_{LP} \label{eq:G}
\end{eqnarray}

{\color{black}{Eq. \ref{mass} expresses that since matter cannot disappear, the total amount of a given resource is constant and equal to the initial value $X_T$ present throughout the system. The flow of resource input $F_{HP}$ is assumed to depend on the
high level potential $\mu_H$ and on intensity $J_P$.}}

If $\mu_H$ is high, then access to resource is easy and the flow of resources, $F_{HP}$, must be large for high levels of intensity, cf. Eq.\eqref{eq:FHP}. The potential $\mu_H$ does not influence the quantity of waste. On the other hand, if intensity $J_P$ is large, the amount of produced waste will also be large. This is expressed by Eq.\,\eqref{eq:FLP}. Moreover, we need to take into account the fact that, if the conversion engine operates at a higher intensity, $J_P$, the amount of wastes will increase more than linearly. Hence, the quadratic element in\,\eqref{eq:FLP}. The last two equations are again a consequence of mass and energy conservation: the mass flux of produced goods is the
difference between the input and output flows, see Eq.\,\eqref{eq:G}. The flow of goods can either directly provide the manufacturers or, if there is an excess of production over demand, be stored in the buffer zone, see Eq.\,\eqref{eq:FLP} and Eq.\,\eqref{eq:FHP}. Conversely, if aggregate demand exceeds supply, the difference is taken from the stock of inventories, when possible.

One can notice that $G$ is a parabolic function of $J_P$ presenting a maximum at $J_{P}^\text{max}=(\mu_H - \mu_L)/2 \, R_p$ and a zero production at $\overline{J}_p=2 \, J_{P}^\text{max}$. That no intensity higher than $\overline{J}$ enables the economic metabolism to produce positive work is a generalization to our E\&M framework of the idea that an EROI lower than 1 does not enable to extract any energy. That $I
^*_p$ is finite means that maximising {\sc gdp} growth (as a monetary proxy for maximising $J$) does {\it not} enable to maximise useful work.

Whenever demand exceeds supply and the buffer stock is empty, then aggregate demand is rationed, so that  the fulfilled part $G_D^\text{satisfied}$, or $G_D^S$ in the following, of demand reads:
\begin{eqnarray}
G_D^\text{S}:=&
    \begin{cases}
    	G_D	& 	\begin{cases}
    			\text{ if } G-G_D>0 \\ \text{ or } X_S>0
		    	\end{cases}\\ 
	    G   & \text{ otherwise. }
    \end{cases}
\label{eq:GDS}
\end{eqnarray}

To close the system, we also need to express the potentials with respect to other variables. The higher the stock, $X_H$ of resources,
the higher the potential $\mu_H$;  the same holds for  $X_L$ and $\mu_L$. For instance, one can adopt
\begin{eqnarray}
\mu_H &:=& \tanh \left(\alpha  \, \frac{X_H}{X_T}\right)\label{eqnpotH}\\ 
\mu_L &:=&  \tanh \left(\alpha  \, \frac{X_L}{X_T}\right)\label{eqnpotL}
\end{eqnarray}

Initial conditions (including $0<\alpha\leqslant 1$) are chosen such that $\mu_H > \mu_L$ holds true throughout the time evolution of  stocks.\footnote{Note that $\alpha=1$ worked for all our considered computations.} The choice of the hyperbolic tangent stems from the fact that $\mu_K$ (with $K\equiv H$ or $L$) must first increase linearly with $X_K$, and then saturate from above. It is therefore clear that these two potentials are not directly intended to account for the thermochemistry of recycling, but are rather a probe of the state of resources in the economic sense of the term. They are asked to provide a signal on the state of resources and wastes and make it possible to estimate the cost of recycling the said resource. Note that, we considered that no direct substitutability is possible between used resources. 
Therefore, a resource is used until depletion, while filling the secondary residuals sink. System operation can hence be constrained either by resource exhaustion or by too much waste.\footnote{This is consistent with thermodymanics, as previously stated in Section\,\ref{Conv}).} In order for the economic metabolism to be able to produce, we need to add a recycling process, efficient enough to regenerate resources and drain the waste reservoir. This will be represented by the two right hand side blocks of Fig.\,\ref{schemacomplet} and described in section \ref{ssec:recycling}.

\subsection{Potentials, Entropy and Exergy}

The ontological issue regarding the relationships between thermodynamics and economics is central\footnote{The general question of the relations between thermodynamics and economics is old and well documented, and our purpose here is not to revisit the debates that have already taken place. Readers interested in these questions will find elements of ontological questioning in the work of Spash \cite{Spash2012}. \cite{Maki2009} also offers a complementary look that addresses the issue of scientific imperialism in this debate.} in that, while matter exists in its physical and economic definitions respectively, this is by no means sufficient to conclude that both disciplines share a common ontology: the capacity of the economy to account for matter in monetized form, and its difficulty to do it in physical form\footnote{At least, without postulating some kind of substitutability between matter, money and/or labour.} suggests the presence of multiple ontologies.

In this section, we would like to clarify a few points concerning the thermodynamic measurement of resource quality.
In thermodynamics, the entropic principle stipulates that, in addition to the measurement of quantities, one can also measure qualities. Thus, energy and matter are given two characterizations, whereas all other disciplines of physics (mechanical, electromagnetic...) only consider quantities. This is well illustrated by Carnot's yield, which stipulates that efficiency of a heat engine is necessarily lower than Carnot's factor $(1-T_c/T_h)$, where $T_c$ and $T_h$ are the cold and hot tank temperatures respectively. The central issue, which alone sums up the singularity of thermodynamics, is that of the gauge. Indeed in most physical disciplines the choice of the gauge, \textit{i.e.} the zero of the scale, is  arbitrary. This is the case in electrokinetics with the choice of the ground potential, or in mechanics with the choice of a zero of the altitudes at the sea level. Things are radically different in thermodynamics where the gauge is rigidly defined in the temperature scale called thermodynamic temperature. If it were not so, it would be possible to arbitrarily define the following gauge $T_c=0$ or any other value. We see then that with the choice $T_c=0$, Carnot's efficiency would be equal to unity: thermodynamics would lose its singularity, and consequently, its very existence. The second principle and the notion of entropy thus boil down to the existence of an indisputable fundamental gauge. The concept of exergy then immediately follows from Carnot's factor: it is only a dual representation of the second principle, in which the fraction of energy really usable in the form of work ---also called free energy--- is the total available energy multiplied by Carnot's factor. However, and for practical reasons, the exergy approach often uses a chosen reference temperature, which amounts to defining an arbitrary gauge. As a consequence, the concept of exergy is only a weak version of the second principle, whereas the entropy concept is a strong version of the second law of thermodynamics.

 If the question of the energy gauge is rather simple because it is based on the very definition of what the thermodynamic temperature is, it is not the same for the gauge which concerns material resources. As already said, the second principle also applies: like energy, matter is subject to dispersion. The measure of an element of matter in the sense of its environment is defined by its chemical potential.  This does not, however, give access to an absolute gauge for matter, \textit{i.e.} an absolute zero of chemical potentials. An elementary reasoning supported by an equally elementary thermodynamic calculation shows that the energy cost for the concentration of a diluted resource diverges as a logarithm function with the dilution rate.  The problem posed is therefore that of the existence of a gauge that would characterize the ultimate dilution, \textit{i.e.} the disappearance of matter. Because of the first law, no absolute gauge can therefore qualify matter in a strong version of the second law,  such as the one that exists for temperature. Whereas there is indeed a scale of absolute temperatures, there is no absolute scale of chemical potentials. If, now, a weak version of the second law is accepted, an energy-efficient approach to material qualification can be considered. This  approach has been widely developed in the last twenty years by Valero, and before him by Ayres, who first introduced the concept of exergy in the field of ecological economics. In order to realistically situate the problem at the scale of our Earth, which is a closed material system, the choice of gauge is usually defined by the ultimate concentrations that can be found in the Earth's crust, oceans or atmosphere. This viewpoint allows, via a standard thermodynamic calculation, to evaluate the energy necessary to the concentration of a chemical element thus diluted, until reaching its initial ore concentration. Such a strategy may seem quite realistic in that it seems to be in perfect agreement with the laws of thermodynamics. 
 
 However, three criticisms can be formulated. The first one, already developed, concerns the weak form of the second law that represents the exergetic approach. In more practical terms this means that as soon as the gauge becomes an adjustable parameter, the exact relation to thermodynamics weakens strongly to the point that to assert that the energetic approach to the treatment of matter is more realistic than another becomes questionable. The second criticism comes from the numerical estimates that emerge from these calculations and their economic validity. Indeed, the energy cost of recycling a diluted resource becomes dependent on the choice of gauge, which reduces the relevance of the estimate not better than to an order of magnitude. This criticism can be modulated by the fact that, since the dependence is logarithmic, the sensitivity to the choice of gauge is limited for a wide range of concentrations, except for concentrations close to the minimal value set by the gauge. It should be added that while the energy cost of concentrating a resource can be estimated, nothing is said about the quality of the energy required for this operation. Indeed, in certain thermodynamic favourable situations, simple fusion-aggregation may suffice, which amounts to using only heat. In other cases, the input of work will be indispensable, as in the case of osmotic separation processes for example. To mention only the ecological field, it can be noted that the sustainability of living organisms results from the supply of very high quality energy in the form of solar photons, and that a simple supply of heat in equal quantities would not lead to the closing of the cycle of living organisms via the photosynthetic process. As everywhere in thermodynamics, it is through a high quality energy input that the cycling processes, which are counter-entropic by definition, are carried out.
Adding this quality dimension to the discussion, we can see that the strictly thermodynamic exergetic approach does not allow an economic estimate as well founded as a first reading would suggest.
This is why we have not chosen a strictly thermodynamic formulation of resource potentials, so that the analytic expressions of the potentials are  given in normalized form in Eqs.\,\eqref{eqnpotH} and\,\eqref{eqnpotL}. 
Note that, although these definitions do not correspond analytically to that of a chemical potential, they nevertheless define two potentials, and, in doing so, make it possible to consider at the same time $(i)$ an assessment of the available resource, $(ii)$ an assessment of the unavailable resource, $(iii)$ a measure of the criticality of supply of the resource, $(iv)$ a criterion for pinching production due to the accumulation of the unavailable resource, and therefore the need for recycling.

Let us now come back to the difference in potentials, $\Delta \mu=\mu_H-\mu_L$, which acts as a driving force for the conversion process.
This gradient is the primary cause that makes it possible to convert E\&M, whether in manufacturing production, in services or in the production of living organisms. In living organisms, this is maintained by the constant flow of solar energy, which, being of high quality, makes it possible to maintain a difference in chemical potential throughout the trophic chain. If this flux was to cease, the matter located in the structures, especially living ones, would inexorably end up dispersing. This is similar to what is encountered in photovoltaic conversion \cite{alicki_thermodynamic_2017}. 
Observe as well that the high quality of solar energy is at the origin of natural recycling. This is an invariant feature of thermodynamics: lowering the entropy of a system requires a very high quality energy input. This of course applies beyond the case of natural recycling, and a manufacturing production can only become cyclical provided it receives an influx of high quality energy, ensuring forced recycling. This is necessary to guarantee that the potential $\Delta \mu$ difference does not vanish. Indeed, the eventual pinching over time of $\Delta \mu$ is a main determinant of a crisis in the production process.

Of course, and for a limited time period, fossil fuels may contribute to recover this $\Delta \mu$ difference, in a forced recycling process but, as we shall see shortly,  the drawbacks are important.

In each sheet we then get a full blown description of consumption and recycling of resources by knowing the quantities $X_k$ and the qualities $\mu_k$.
The formulation of the exergy factor $E_x=\frac{\Delta \mu}{\mu_H}$ is then straightforward. Moreover, the force and flux approach allows to directly consider the entropy production. According to \cite{onsager1931}, in a close to equilibrium thermodynamic derivation, the production of entropy is given by the product of the force by the flux. The balance of the incoming and outgoing entropy flows during the production process in a sheet is therefore simply given by
\begin{eqnarray}
    \bigdot{S}&=& \frac{F_{LP}}{\mu_L}-\frac{F_{HP}}{\mu_H}= \frac{R_P \, J_P^2}{\mu_L}
\end{eqnarray}
For a reversible production process, one would get $\bigdot{S}=0$ so $\frac{F_H}{\mu_H}=\frac{F_L}{\mu_L}$, or alternatively, $R_P\simeq0$. 
We would like to underline, here, that minimizing the $R_P$ impedance can be understood as the means by which an energy conversion system maximizes the convective transport of matter. As recalled in \cite{goupil2020}, this convective transport is at the very origin of the conversion of energy.\footnote{When the biosphere is  understood as a meta-system of energy conversion, it is sometimes suggested that the operating point must maximize power produced, erected as a  principle of maximum MPP power \cite{Hall}. In addition, it is also empirically observed hat there is a strong tendency for systems to move towards an intensification of the free energy density that runs through them \cite{chaisson_energy_2011}. This last statement is not problematic in the sense that free energy flow maximization is the dual form of convective flow maximization, i.e., minimization of $R_P$. On the other hand, to claim that this situation necessarily corresponds to a MPP is questionable, see \cite{goupil_thermodynamics_2019}.}
This first observation on $R_P\simeq0$ might seem rather partial because it concerns only one variable, $R_P$. However, each sheet has its own variable, $R_P$, and a concrete economy is made up of millions of such sheets. The global minimization of the entropy production on the scale of an economic system thus emerges from complicated trade-offs between the values of $R_P$ within the respective sheets.

Since there is no such thing as self-organization without high quality energy input, this requires an additional extraction of energy from the reservoir. 
Considering a unique sheet we get the following equations in addition to Eq. \eqref{mass}-\eqref{eq:G}:
\begin{eqnarray}
\bigdot{E}_{HP}          &=& F_{HP} \; \left(1-\frac{\mu _{L}}{\mu _{H}}\right) =\Delta \mu \, J_{P} \\
\bigdot{E}_{LP}          &=& F_{LP} \; \left(1-\frac{\mu _{L}}{\mu _{L}}\right) =0 \\
\Delta \bigdot{E}_{X}    &=& \bigdot{E}_{HP}-\bigdot{E}_{LP} 	=\Delta \mu \, J_{P}	\\
\eta            &=& \frac{G}{F_{HP}} 					\\
\varepsilon     &=& \frac{G}{E_{HP}}	  
\end{eqnarray}

$\bigdot{E}_{HP}$ is the incomming exergy flux, $\eta$ is the energy  efficiency and $\varepsilon$ the exergy efficiency. Notice that $\bigdot{E}_{LP}=0$ which simply means that what is rejected at this level no longer has any exergy value.
Some expressions are of specific interest. The first one is the maximal production $G_{\text{max} } = \frac{\left( \Delta \mu \right) ^{2}}{4 \, R_{p}}$ where the intensity at maximal production is $ J_{P}^\text{max}=\frac{\Delta \mu }{2 \, R_{p}}$. Also important are the entropy fluxes and the entropy production $\bigdot{S}$,  
\begin{eqnarray}
\bigdot{S}_{HP}    &=&\frac{F_{HP}}{\mu _{H}}   = J_{P} \\
\bigdot{S}_{LP}    &=&\frac{F_{LP}}{\mu _{L}}   = J_{P} + \frac{R_{p} \, J_{P}^{2}}{\mu _{L}} \\
\bigdot{S}         &=& \bigdot{S}_{LP}-\bigdot{S}_{HP} = \frac{R_{p} \, J_{P}^{2}}{\mu _{L}} 
\end{eqnarray}

As already said, the production, $G$, follows a parabola curve from $G(J_P=0)=0$ to $G(J_{P}=\frac{\Delta \mu}{R_P}) = 0$ which is a short circuit configuration. According to a general thermodynamic approach \cite{goupil2020}, we can identify three characteristic intensities in this curve. Increasing $J_P$, we first reach the point of maximal energy efficiency.

Then, we reach the maximal production (maximal power) point already mentioned. It is worth noticing that beyond  $G_\text{max}$ both production and efficiency decrease. This is clearly a detrimental configuration. 
Let us go back to the exergy analysis. The exergy efficiency, $\varepsilon$, may be rewriten as 
\begin{eqnarray}
\varepsilon  &=&\frac{G}{\bigdot{E}_{HP}}=1-\frac{J_{P}}{2 \, J_{P}^\text{max}}.
\label{varepsilon}
\end{eqnarray}

The variable $\varepsilon$ is, of course, neither constant across time nor identical across countries. Its mean, however, seems to have been orbiting around $\varepsilon \sim 0.19$ in the early 2000s' \cite{ayres_economic_2010}). 
According to Eq.\,\eqref{varepsilon} we get the value of the average intensity of the production $J_{P}=1.62\times J_{p }^{\text{max}}$. 
This observation suggests hat the system is working well beyond its maximal production zone, in a detrimental region where both production and efficiency are far from being maximal.

According to \cite{ayres_economic_2010}, the ratio between useful work and real {\sc gdp} for the US, UK and Japan were remarkably close to each other in 2000, around $\sim$1.6 MJ/US\$. Despite the idiosyncratic stories of each of these countries, all these country-specific ratios where declining since their peak around 1970 at a pace that should have led them close to 1.5\,MJ/US\$ ten years later, provided their dynamics remained identical. In 2010, the world {\sc gdp} was US$\$$ 65,96 trillion. Extrapolating $\eta\sim  1.5$ for the entire world, an educated guess leads to ${\mathcal W}\sim 98.33\times 10^{18}$J, hence 2.35 Gtoe in 2010. Now, we need to take account of the fact that, by construction, the {\sc gdp} metric neglects the informal sector ---which obviously also relies on energy and matter. The informal sector (including the underground economy involving illicit activities like prostitution and the sale of illegal drugs and weapons) is estimated to represent 8-10\% of the {\sc gdp} in the US, and more than 70\% in Ivory Coast. More generally, informal transactions are often estimated to account for one-third of the total economy in developing countries and slightly more than 10\% of the total economy in advanced countries. Let us assume that it roughly amounts to 25\% of the total {\sc gdp} at the world level. We therefore get an estimation of $G\sim 122.91\times 10^{18}$\,J for the world economy, in 2010.

\subsection{Recycling}
	\label{ssec:recycling}

Let us now focus on the recycling zone inside a sheet. Once the resource is used and partly transformed into secondary residuals, it can be recycled either $(i)$ through the forced recycling  zone that runs like the production zone (the central zone of Fig.\,\ref{schemacomplet}, with the engine $R_R\, J_{R}$) or $(ii)$ through natural recycling (depicted by the arrow with flux $F_{NR}$ at the right hand side of Fig.\,\ref{schemacomplet}). Natural recycling mainly concerns primary biomass, but may also correspond to reproduction for animals, plants, bacteria... 
Forced recycling mainly deals with wastes unable to recycle endogenously on their own and that require human technical assistance to be re-transformed into resource. 
Note that a continuous recycling of resources assumes perfect recyclability, which is a strong assumption. We also consider that the recycled resource 
possess the same quality as that of the pristine resource in its primary state \cite{Ayres1969}. Real recycling cannot be perfect neither in quantity nor in quality.
For simplicity, let us assume that we are able to perfectly recycle the resource both in quality and quantity\,:\footnote{This is clearly not true for many metals for instance.} if we were able to recycle the whole secondary residual, no resource rarefaction would ever appear. For the forced recycling zone, the system of balance equations is  similar to system\,\eqref{Intensityprod}--\eqref{eq:G}, of the production zone:
\begin{eqnarray}
J_{R}    	&=& n_R  \, J \label{intensityRecycling}\\
F_{HR} 	&=& \mu_H  \, {J_{R}} \label{eqnR4}\\
F_{LR}	&=& \mu_L  \, J_{R} - R_{R}  \, {J_{R}}^2 \label{eqnR5}\\
F_{RIn}	&=& F_{HR} - F_{LR} \label{eqnR6}
\end{eqnarray}

In order to make the recycling engine work, a flow of primary resource, $F_{RIn}$, is required. $F_{LR}$ denotes the input flow of residuals
into the recycling engine. A flow of recycled resource $F_{HR}$ is then injected back to the high potential resource zone. Intensity, $J_{R}$, in Eq.\,\eqref{intensityRecycling} and friction coefficient, $R_{R}$, respectively parallel  the $J_{P}$ and  $R_{P}$ parameters of the production zone. All the flows under scrutiny are assumed to be positive, and we define  $J_{R}^{\text{max}} $, see Eq.\,\eqref{eqn8}, as the maximal possible intensity of the recycling zone in order to get a  maximal $F_{LR}$, Eq.\,\eqref{eqnR5}: 
\begin{eqnarray}
    J_{R}^{\text{max}} &:=& \frac{\mu_L}{2 \,  R_{R}} \label{eqn8}
\end{eqnarray}

Obviously, intensity is an essential parameter since it defines the rate at which resources are used.If intensity is too high, the resource might not have enough time to regenerate, and the whole production system may collapse (an example of such a behavior is given in section\,\ref{Example}). The natural recycling flow, $F_{NR}$, features the renewal of a resource by natural processes. It represents, for example, reproduction in the case of biomass. In the literature, the natural rate of growth is governed by the very classical Verhulst logistic equation, $ F_{NR} = r \,  X_H  \, (1 - T_H)$ with $\displaystyle{T_H= \frac{X_H}{X_T}}$ and $r$ being the regeneration rate.
Several ecological studies \cite{Liermann} point out that, for many species, the fertility rate, $r$, decreases when the population is excessively reduced. This phenomenon is known as the Allee effect \cite{BIOMAT}:
\begin{eqnarray}
F_{NR} &=& r \; X_H \; \left( 1 - T_H \right) \; \left( \frac{T_H}{s} -1 \right) \label{FNR}
\end{eqnarray}

In Eq.\,\eqref{FNR}, the carrying capacity is the initial and maximum quantity of initial resource, $X_T$, and $s\in [0,1]$ is the fraction of this carrying capacity representing the threshold below which $r$ decreases.
In the next subsection, we shall now write the governing equations for the stocks dynamics.\footnote{A inflow of resource in a stock is positive while an outflow is negative.}   

\subsection{Stock balance equations}
The  balance differential equations for the high and low stock variables, $X_H$ and $X_L$, are\,:
\footnote{Put together, \eqref{eq:GDS} and \eqref{eq:dotXL} are valid as long as the buffer stock remains positive, so that $G_D$ is never rationed. This is always the case in the numerical simulations to follow, and is a standard short-cut which saves us from considering rationing schemes (see, \textit{e.g.} \cite{GRASSELLI20181}. Doing otherwise would lead us to define the (possibly rationed) ``satisfied demand'', $G_{D}^{\text{satisfied}} := \min \{F_{HP}, G_D\}$, which would introduce non-differentiable boundary issues in the main vector field under scrutiny. For the sake of simplicity, we avoid these technicalities.}  
\begin{eqnarray}
\bigdot{X_H}	&=& F_{NR} - F_{HP} + F_{HR} - F_{RIn} \label{eqn9}\\
\bigdot{X_L}	&=& - F_{NR} + F_{LP} - F_{LR} + G_{D} \label{eq:dotXL}\\
\bigdot{X}_S	&=& G_S = G - G_D \label{buffer}
\end{eqnarray}  

The variation of the stock of primary resources, $X_H$, is due to the inlet of recycled resources ($F_{NR}$ and $F_{HR}$) and the outlet of resources used for production, $F_{HP}$, or recycling, $F_{RIn}$. The sink stock, $X_L$, is filled when goods/services are produced (with flux $F_{LP}$) and also after the final goods are used. Natural and forced recycling allow to regenerate the resource respectively with fluxes $F_{NR} $ and $F_{LR}$ stemming from the sink stock. Note that the buffer zone, $X_S$, does not produce waste.

\subsection{Sheets and demand function} 
	\label{ssec:centralkernel}

\begin{figure}
	\begin{center}
		\includegraphics[width=1\linewidth]{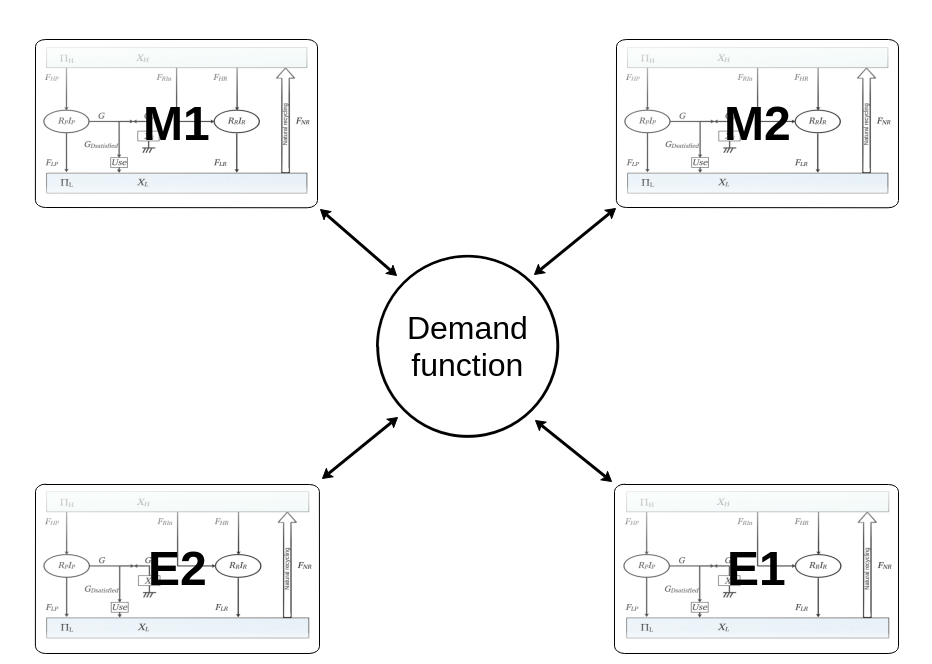}
		\caption{Physical sheets with four sectors interconnected by a central kernel }
		\label{schema4secteurs}
	\end{center}
\end{figure}

Obviously, the real world is more complex than the single sheet described above. A more realistic model has to take into account many kinds of resources and their interactions. Each sheet, thanks to its structure, can easily describe each resource on Earth. However, interactions between resources is lacking. We propose to represent these complex interactions through a central kernel where no production function exists anymore but a \textit{demand function}, i.e., a mapping  addressing to each sheet a request for resources. The demand function coincide with a production function only if resources are infinite. The economy is modeled by multiple sectors or resources (whose dynamics are governed by their elementary sheets) that can be all connected to a central kernel.
The kernel addresses requests (the \textit{demand function})  to these sectors (the sheets), that may be able to satisfy them or not. 
Fig.\,\ref{schema4secteurs} presents an example of a 4-sector economy (two material resources M1 and M2 and two energy resources E1 and E2)
interconnected through a central kernel.

\section{Resource Case studies}
\label{casestudy}

\begin{center}
	\begin{tabular}{|c|c|c|}
		\hline 
		Symbol	
			& Description 
				& Value \\ 
		\hline 
		$X_T$	
			& Total quantity of resource 	
				& 1000.0 \\ 
		\hline 
		$r$		
			& Regeneration rate 
				& 0.025 \\ 
		\hline 
		$s $		
			& Threshold percentage 
				& 0.2 \\ 
		\hline 
		$R_{P} $
			& Production friction 
				& 0.001 \\ 
		\hline 
		$R_{R}$	
			& Recycling friction
				& 0.001 \\ 
		\hline 
		$G_D $	
			& ConstantDemand 
				& 30 \\ 
		\hline 
	\end{tabular} 
	\captionof{table}{Parameters of resource case studies}
	\label{tab:param}
\end{center}

In this section, we substantiate our model by several case studies. For the sake of simplicity, we consider only one physical sheet. Initially ($t=0$), the entire amount of resource is available and ready to be converted by the economic engine, while the sink of secondary residuals is empty: $X_H = X_T$, $X_L=0$, $\mu_H =1.0$ (say) and $\mu_L = 0$. At $t=0$, the difference of potential is maximal, as hence is initial production. 

The global intensity $J$ here reduces to   the intensity, $J_P$, of the single sheet.  $J_P$ is chosen such that $G=G_D$ be satisfied by the current aggregate production, \textit{i.e.}, $J_P$ solves the following quadratic equation\,:
\begin{eqnarray}
	- R_{P} \, J_P^2 + \Delta \mu \, J_P - G_D &=& 0 
	\label{Int}
\end{eqnarray}

\begin{center}
    \begin{tabular}{|l|m{1.8cm}|m{3.0cm}|}
	\hline 
	\centering Case/Color  
		& Name 
			& Equation \\ 
	\hline 
	\centering 1 / Blue 
		& Maximum intensity 
			& $J_{P}^\text{max}= \frac{\Delta \mu }{2 \,  R_{P} }$\\ 
	\hline 
	\centering 2 / Green 
		& Optimum intensity 
			& Lower positive root of \eqref{Int} \\ 
	\hline 
	\centering 3 / Red
		& Part of optimum intensity 
			& 20 \% of lower positive root of \eqref{Int}  \\ 
	\hline 
    \end{tabular} 
    \captionof{table}{Intensity for each of the three resource case studies scenarios.}
    \label{tab:cases}
\end{center}

\begin{figure}
	\begin{center}
		\includegraphics[width=1\linewidth]{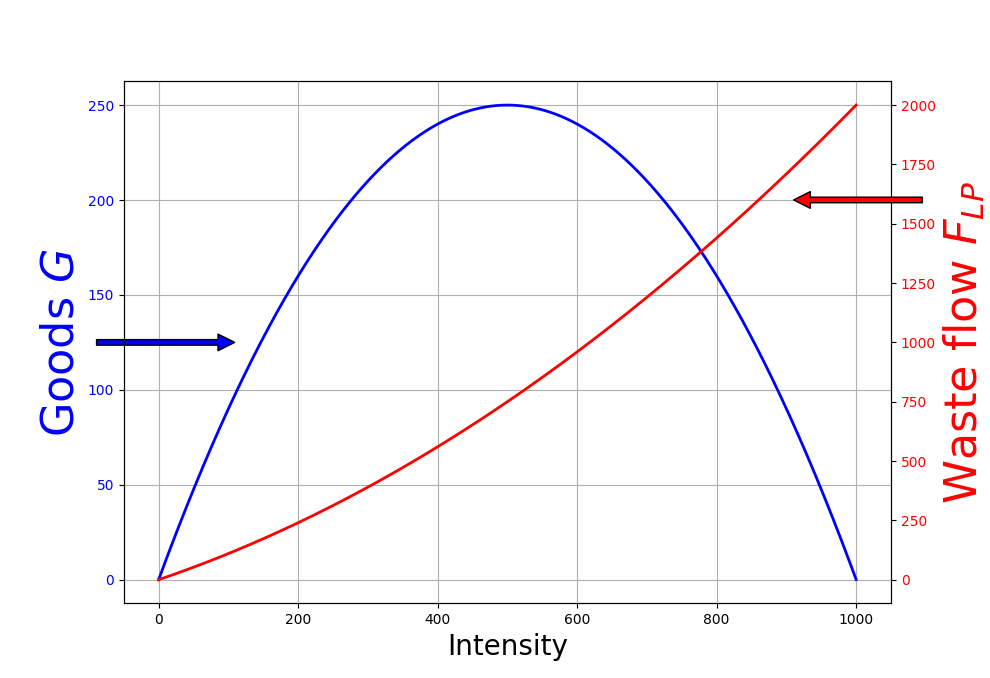}
		\captionof{figure}{Goods and waste flow over intensity, with $R_P=10^{-3}$ and $\Delta \mu=1$} \label{GvsI}
	\end{center}
	\label{fig:puissance}
\end{figure}

Results obtained using Eq.\,\eqref{Int} are illustrated in Figure\,\ref{GvsI}. In order to equate aggregate demand and output, two distinct intensities are available, corresponding to the two roots of Eq.\,\eqref{Int}, except at the top of the parabola. The larger value yields more waste than the lower one. A single level of intensity, $J_{P}^\text{max}$, maximizes current production at $\displaystyle{G_\text{max}= \frac{\Delta \mu^2}{4 \, R_{P}}}$, delivering maximal power, but exhibits less efficiency, as expected:
 \begin{eqnarray}
J_{ P }^\text{ max } &=& \frac{ \Delta \mu }{ 2  \, R_{P} }.
\label{Intmax}
\end{eqnarray}

By contrast, a preferable intensity should satisfy the required demand with minimal waste. For our simulations, we therefore postulate a minimal aggregate rationality by adopting the lower positive root of Eq.\,\eqref{Int} so as to reduce wastes conditionnally on the satisfaction of demand, thus staying below the maximum intensity $J_{P}^\text{max}$ of Eq.\,\eqref{Intmax}. Let us call this an ``optimal intensity'', 
\begin{equation}\label{optimal-intensity}
J_P:=
	\min \, \Bigl\{\text{solution of } \eqref{Int}\; ; \; J_P^{\text{max}}\Bigr\}.
\end{equation}

Of course, whatever being the chosen intensity, the high potential stock decreases over time in every scenario because of dissipation and the low potential stock increases because of waste reservoir filling. The intensity of recycling and the complexity of the metabolism captured through the friction coefficient, $R_P$, however, play a non-trivial role as we shall now see.

\subsection{Intensity impact}
	\label{Intensitystudy}

\begin{figure*}
	\begin{center}
		\begin{tabular}{cc}
			\includegraphics[width=0.5\linewidth]{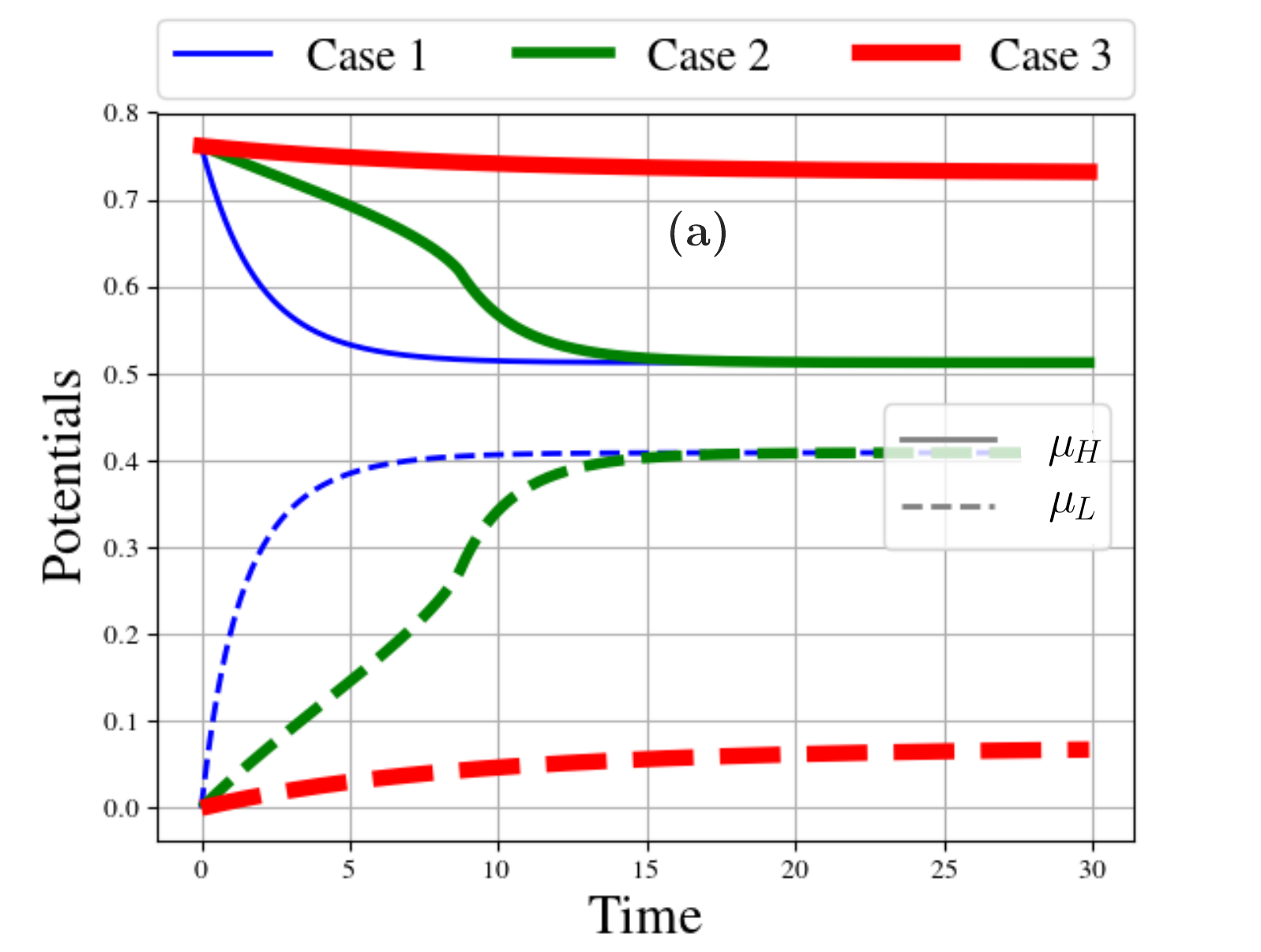}
			\includegraphics[width=0.5\linewidth]{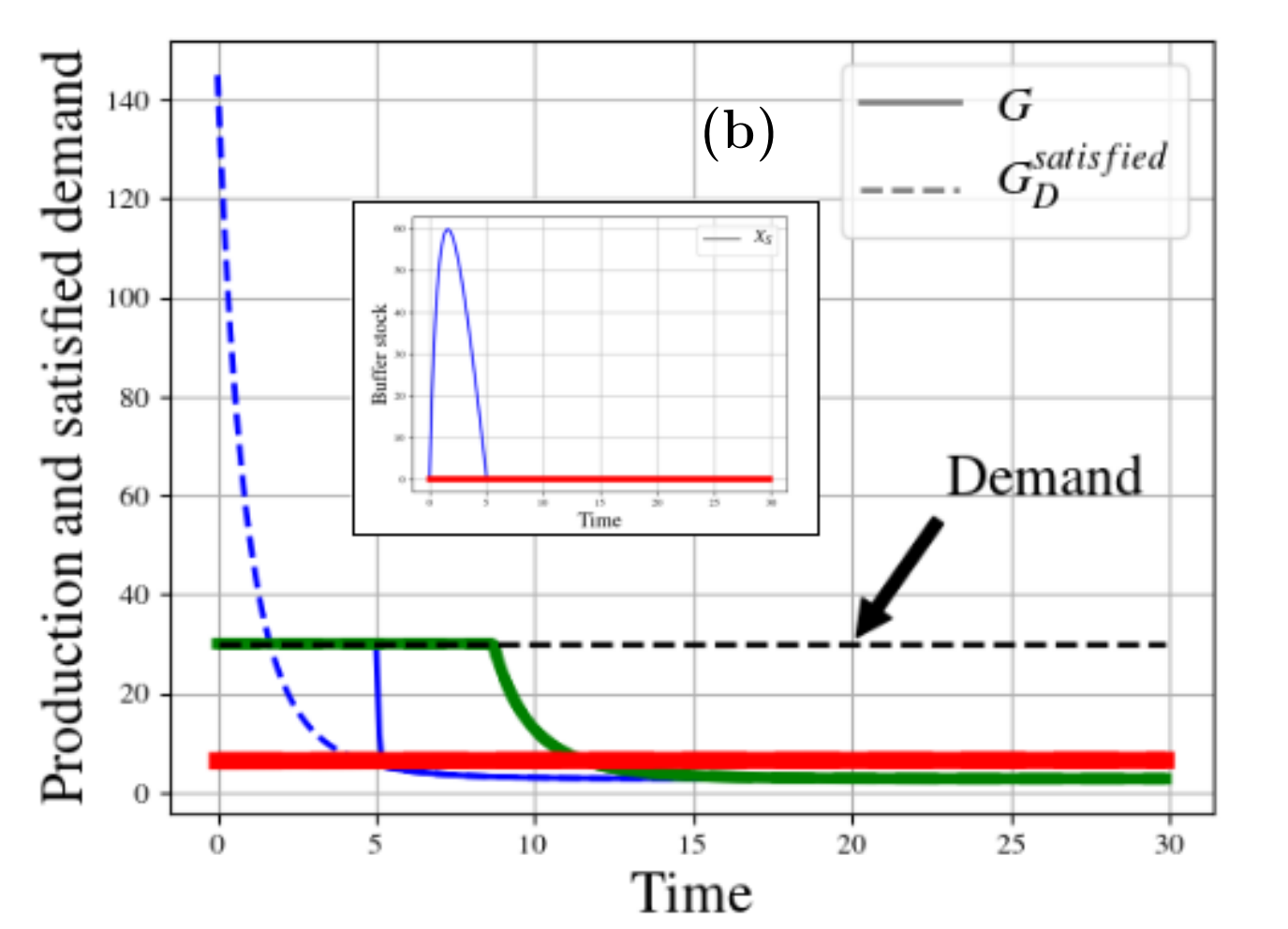} \\
			\includegraphics[width=0.5\linewidth]{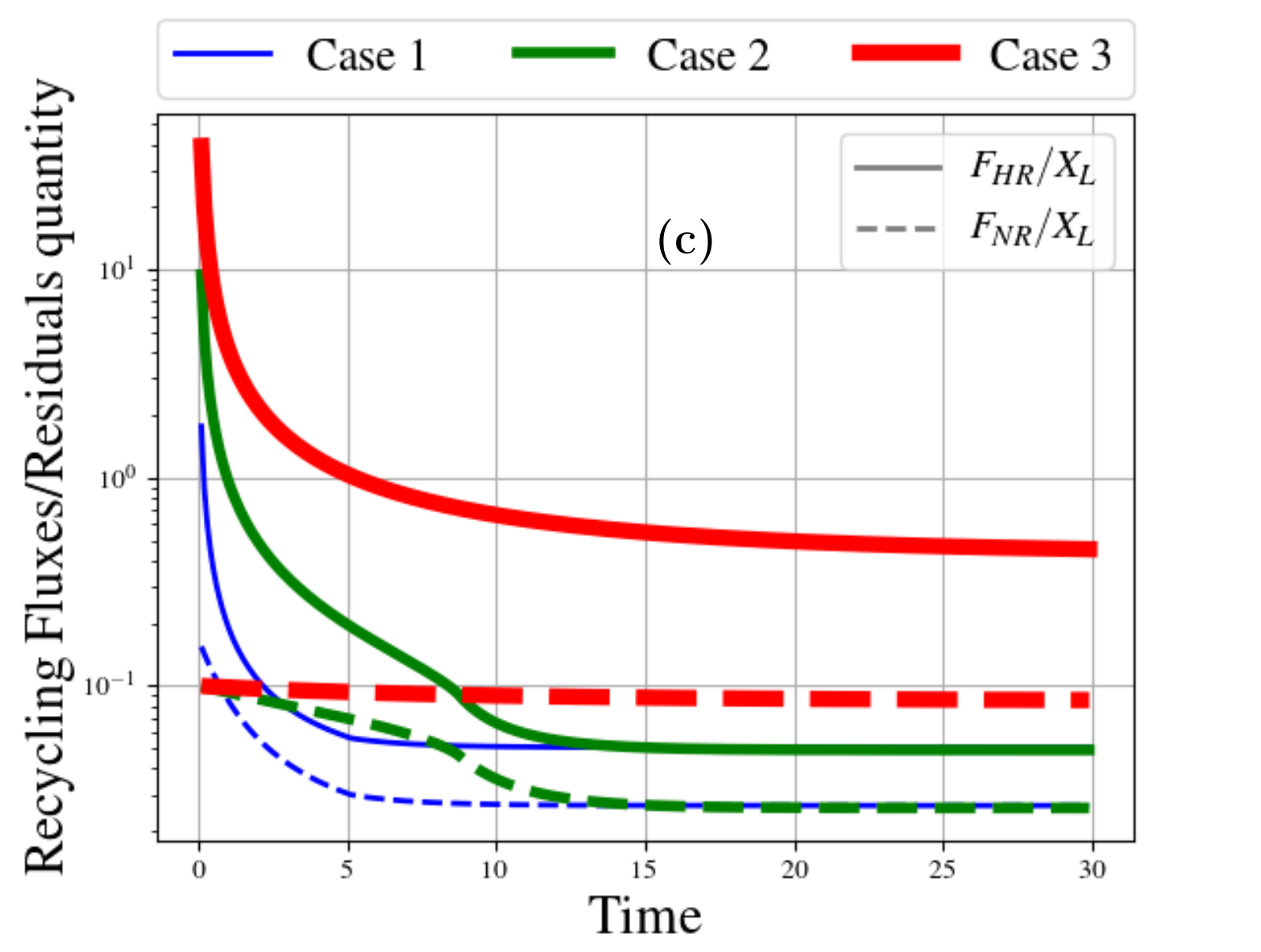}	
			\includegraphics[width=0.5\linewidth]{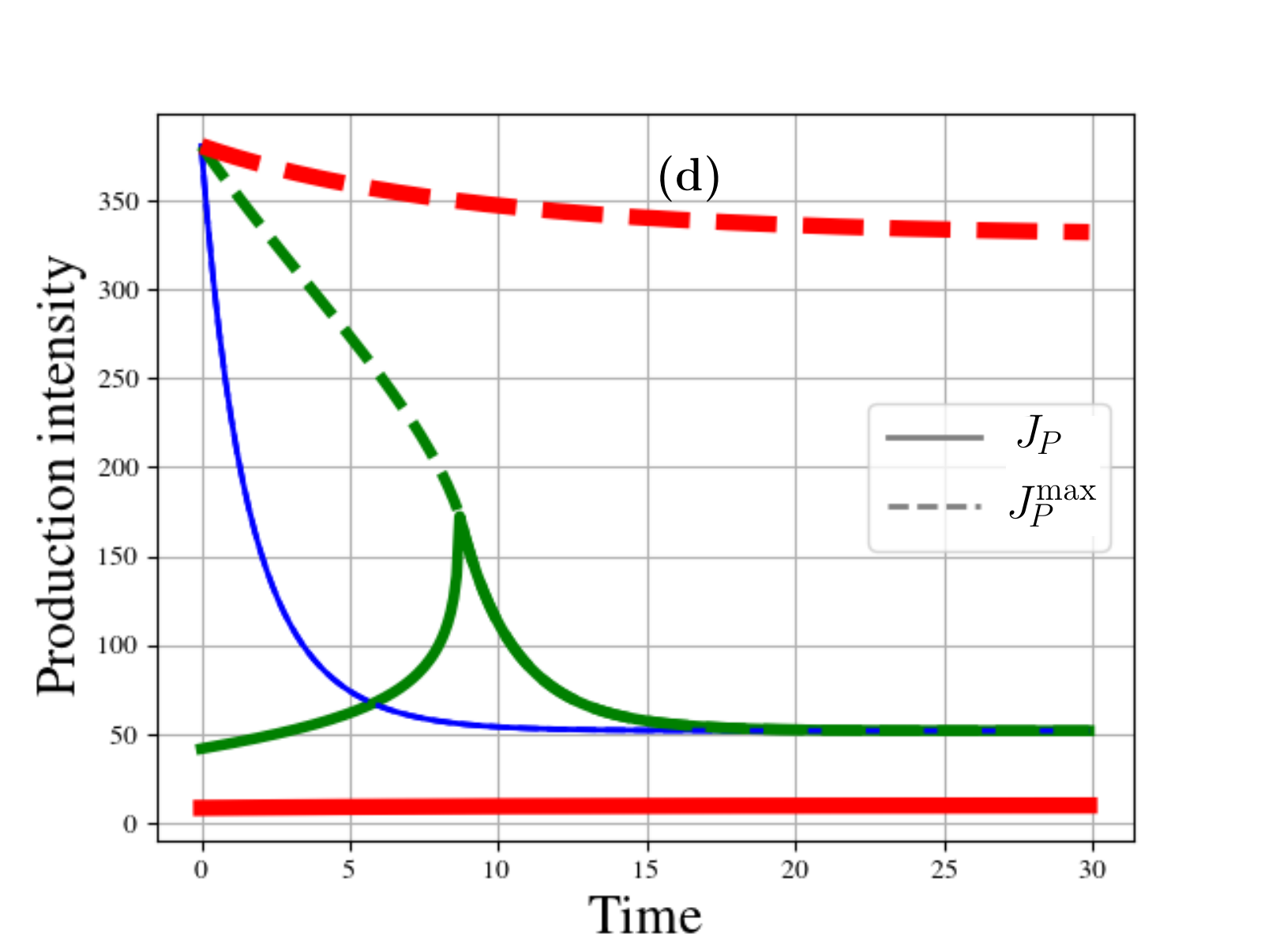}
		\end{tabular}
		\captionof{figure}{Resource case studies: Intensity impact. From left to right and top to bottom. $a$, Time evolution of high and low potentials ($\mu_H$ and $\mu_L$) for maximum (Case 1), optimum (Case 2) and weak intensity (Case 3). The dotted curve is the other positive root of case 1.
			$b$-Time evolution of production of goods and buffer stock for maximum (Case 1), optimum (Case 2) and weak intensity (Case 3).
			$c$-Time evolution of normalized recycling fluxes ($F_{HR}$ and $F_{NR}$) for maximum (Case 1), optimum (Case 2) and weak intensity (Case 3). 
			$d$-Time evolution of intensity for maximum (Case 1), optimum (Case 2) and weak intensity (Case 3)}
		\label{fig:intensityimpact}
	\end{center}
\end{figure*}

We consider three cases, see Tab.\,\ref{tab:cases}: (1) the metabolism runs at the maximal intensity $J_{P}^\text{max}$, (2) it runs at an optimal intensity, and  (3) it runs at 20 \% of this optimal intensity (weak intensity). Furthermore we choose $n_P=1$. In all cases, the demand is kept constant, see Tab.\,\ref{tab:param}.

Figure\,\ref{fig:intensityimpact}-$a$ shows the variations of potentials for these three scenarios. In scenario 1 and 2, potentials end up strongly pinched: the gap between them quickly narrows --- a symptom of the fact that consumption of resources is too fast to be sustainable (Fig.\,\ref{fig:intensityimpact}-b). The pinch-off of the potentials eventually inhibits production, which decreases dramatically as the potentials difference becomes small.

At maximum intensity (case 1), production is initially larger than demand (Fig.\,\ref{fig:intensityimpact}-b), and as consequence the buffer stock (inset of Fig.\,\ref{fig:intensityimpact}-$a$) first fills up. The sheet is able to satisfy the demand at the start (Fig.\,\ref{fig:intensityimpact}-b), when production is high and as long as there are enough goods in the buffer stock ($t\simeq 5$). At optimal intensity (case 2), we succeed in satisfying the demand at the beginning but then, production drops and becomes very low ($t\simeq 10$). On the contrary, at weaker intensity (case 3) potentials are weakly pinched and a large gap remains between the two values but production can never fulfill the demand. It remains nearly constant at a low level.

The curves in Fig.\,\ref{fig:intensityimpact}-$d$ show the evolution of intensities, (solid line), over time, compared to the maximum intensities (dotted line). In the first case the two intensities are of course identical and they collapse due to the rapid pinching of the potentials. In the second case, the intensity increases first of all to meet demand but comes up against the maximum value, so optimum intensity reaches the "maximum" intensity $J_{P}^\text{max}$ when it can no longer meet demand, and therefore follows the slope of the maximum intensity, trying in vain to produce the requested quantity. In the third case the intensity is very low and does not progress although there is a huge margin with the maximum intensity. As for recycling,  comparatively to the total quantity of waste, we recycle a larger fraction $F_{HR}/X_L$ or $F_{NR}/X_L$ in the case of weak intensity (Fig.\,\ref{fig:intensityimpact}-$c$). This is due to the fact that at weak intensity the system runs slow, less resources are used and more time is left to recycle both through natural and forced recycling. At maximal and optimal intensity, waste is produced at a faster rate (Fig.\,\ref{fig:intensityimpact}-$b$) and recycling is not efficient enough. 

\subsection{Recycling impact}
\label{Example}
\begin{figure*}
	\begin{center}
		\begin{tabular}{cc}
			\includegraphics[width=0.5\linewidth]{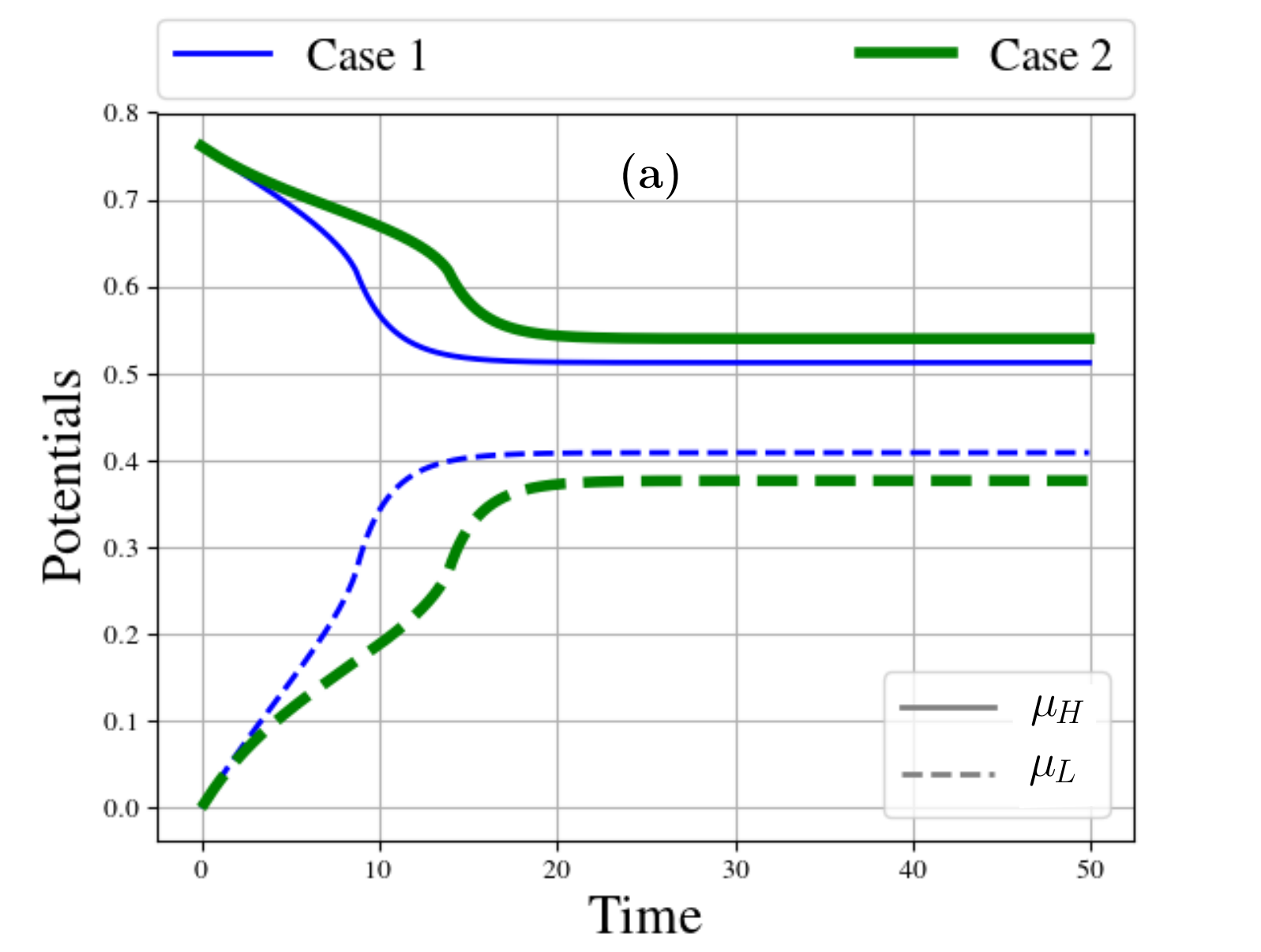}
			\includegraphics[width=0.5\linewidth]{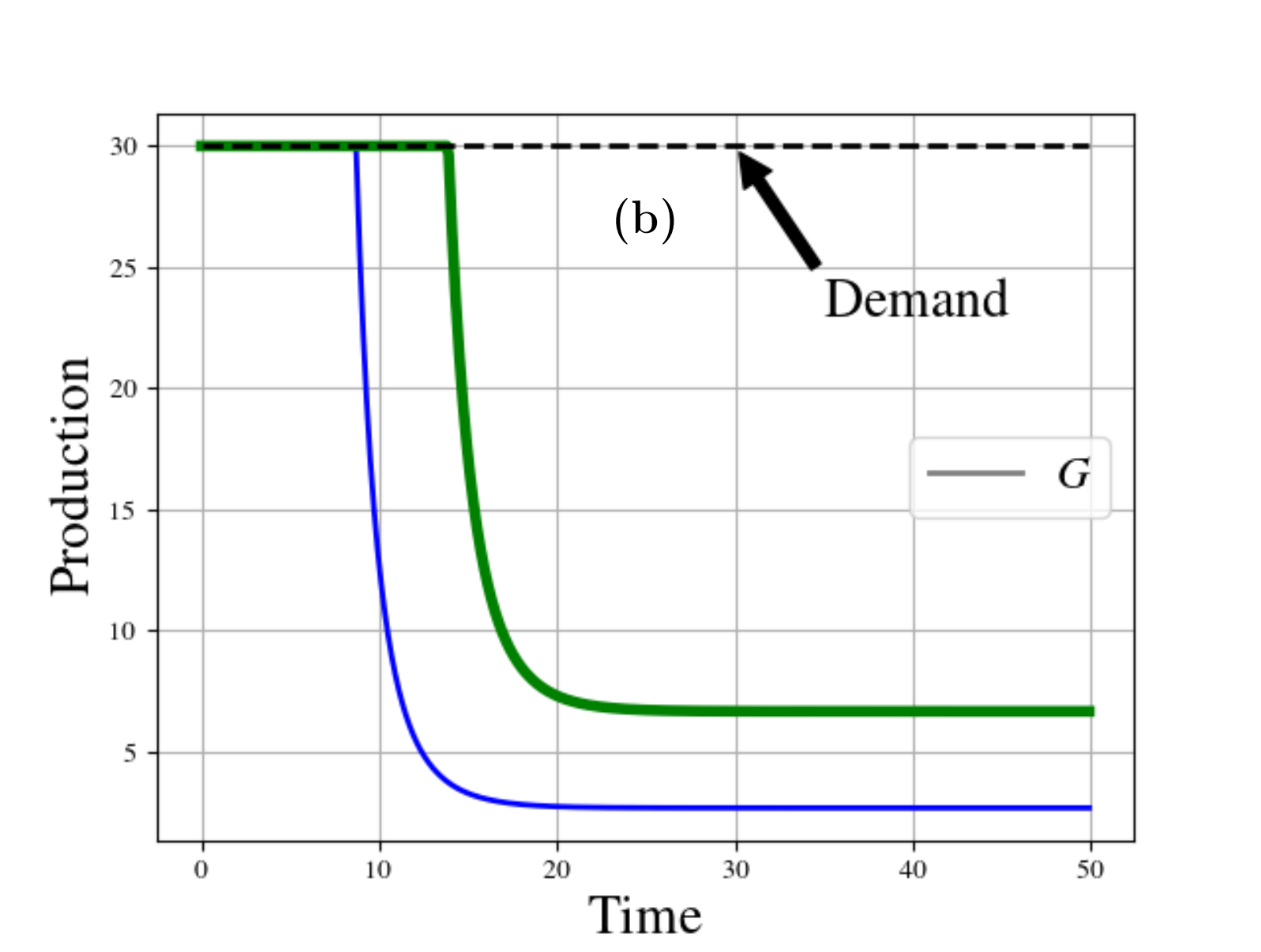} \\
			\includegraphics[width=0.5\linewidth]{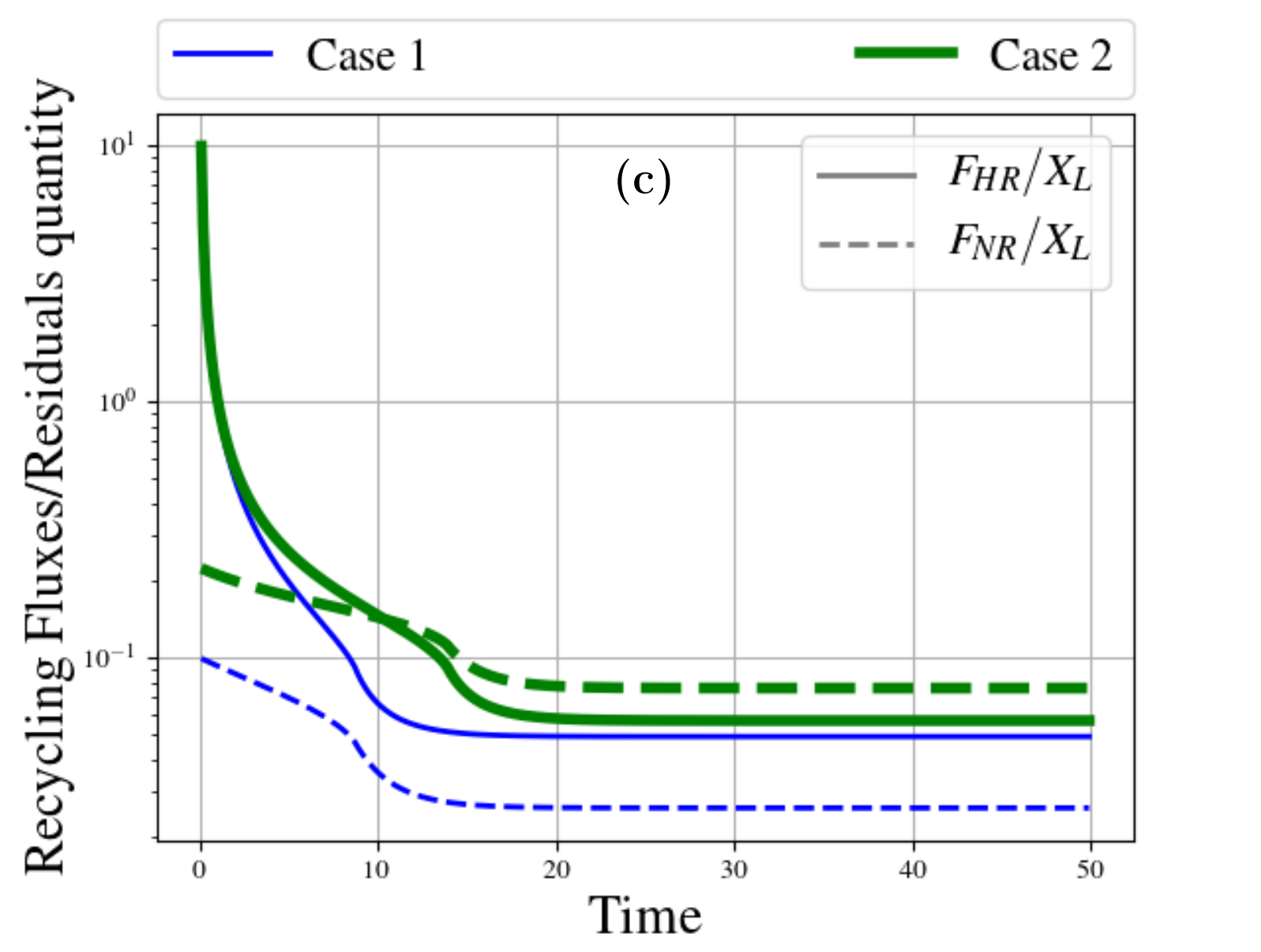}
			\includegraphics[width=0.5\linewidth]{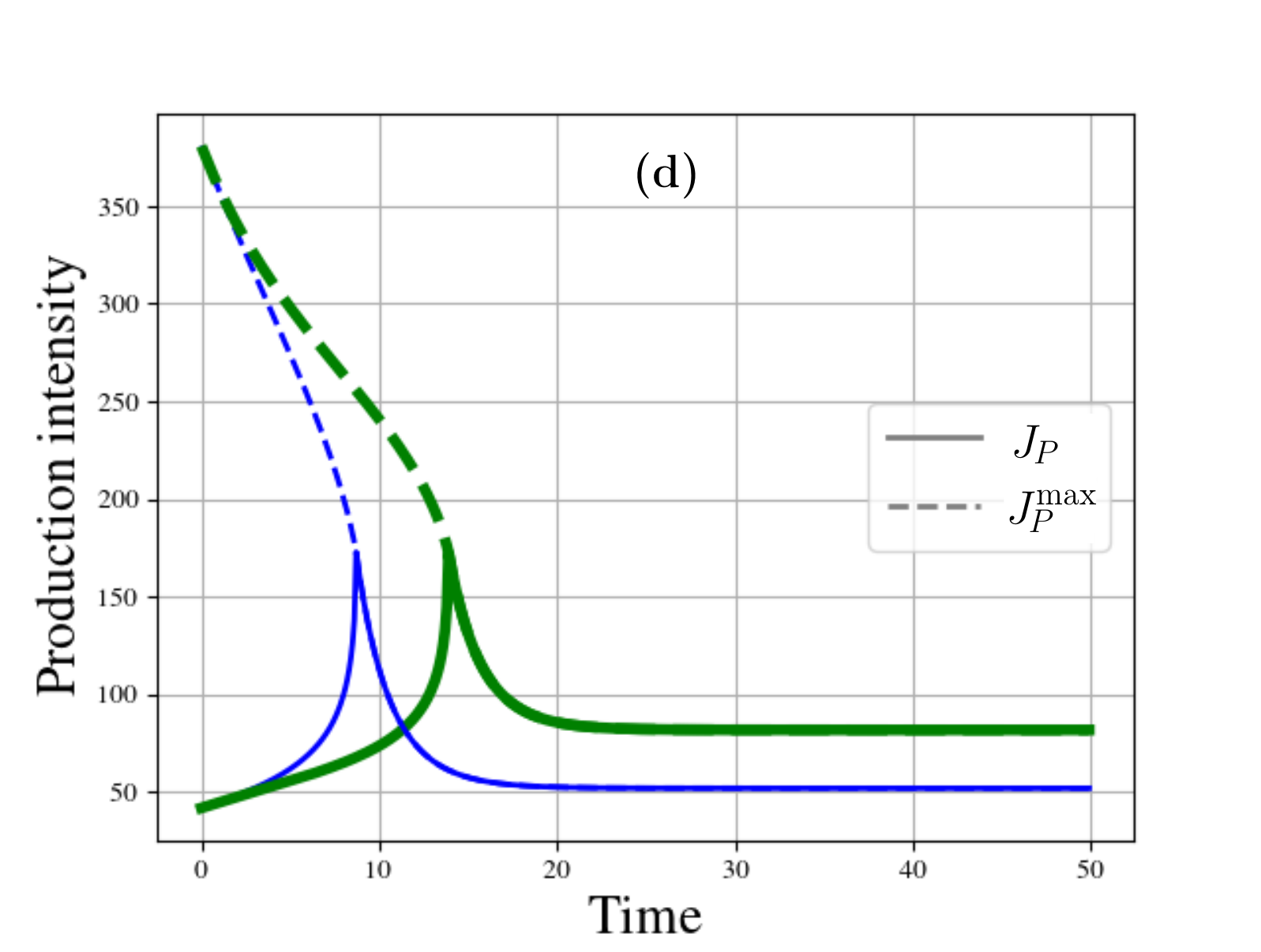}
		\end{tabular}
		\captionof{figure}{Resource case studies: Recycling impact. From left to right and top to bottom. $a$-Time evolution of high and low potentials ($\mu_H$ and $\mu_L$) for high (Case 1) and low threshold (Case 2).
			$b$-Time evolution of production of goods (flux $G$) for high (Case 1) and low threshold (Case 2).
			$c$-Time evolution of normalized recycling fluxes ($F_{HR}$ and $F_{NR}$) for high (Case 1) and low threshold (Case 2). 
			$d$-Time evolution of production intensity (flux $G$) for high (Case 1) and low threshold (Case 2).}
		\label{fig:recyclingimpact}
	\end{center}
\end{figure*}

We study two cases differing by the Allee effect threshold $s$ that affects $F_{NR}$, see Eq.\,\eqref{FNR}. We run at optimum intensity in both cases.
Allee effect threshold $s$ is between 0.1 and 0.2 of the maximum possible quantity for animals \cite{Hutchings}. The regeneration rate $r$ lies in the range $0.01$ --- $0.05$ and we choose an intermediate value $r=0.025$ \cite{Hutchings}. We consider two scenarios. The first one deals with a high threshold at 20\% of the maximum quantity (case 1) and the other with a low threshold at 10\% of the maximum quantity (case 2). That means that in case 1, for $X_H <0.2 \, X_T$, so $T_H/s-1<0$, regeneration is harder and the rate of growth becomes negative. The higher the $s$ percentage is, the earlier the growth rate turns negative. With a high threshold, regeneration is harder. That explains why in case (1) potentials are pinched earlier than in case (2) (see Fig.\,\ref{fig:recyclingimpact}-a). Indeed, natural recycling is lower (see Fig.\,\ref{fig:recyclingimpact}-$c$). 
In both cases, forced recycling rates are close to a common constant rate because the quantities of waste are also close.

The intensity, according to Fig.\,\ref{fig:recyclingimpact}-$d$, reaches the same peak value around 170 for both cases but at different times : case (2) is slower and remains at long times at a constant value larger than case (1). At the beginning, as shown in Fig.\,\ref{fig:recyclingimpact}-$b$, in case (2) production can be satisfied during a longer duration than in case (1). This is because population regenerates at a positive rate during a longer time or during a higher consumption of resource.  Thus, the threshold value $s$ impacts the dynamics as well as the asymptotic values of keyquantities.%

\subsection{Friction impact}

\begin{figure}
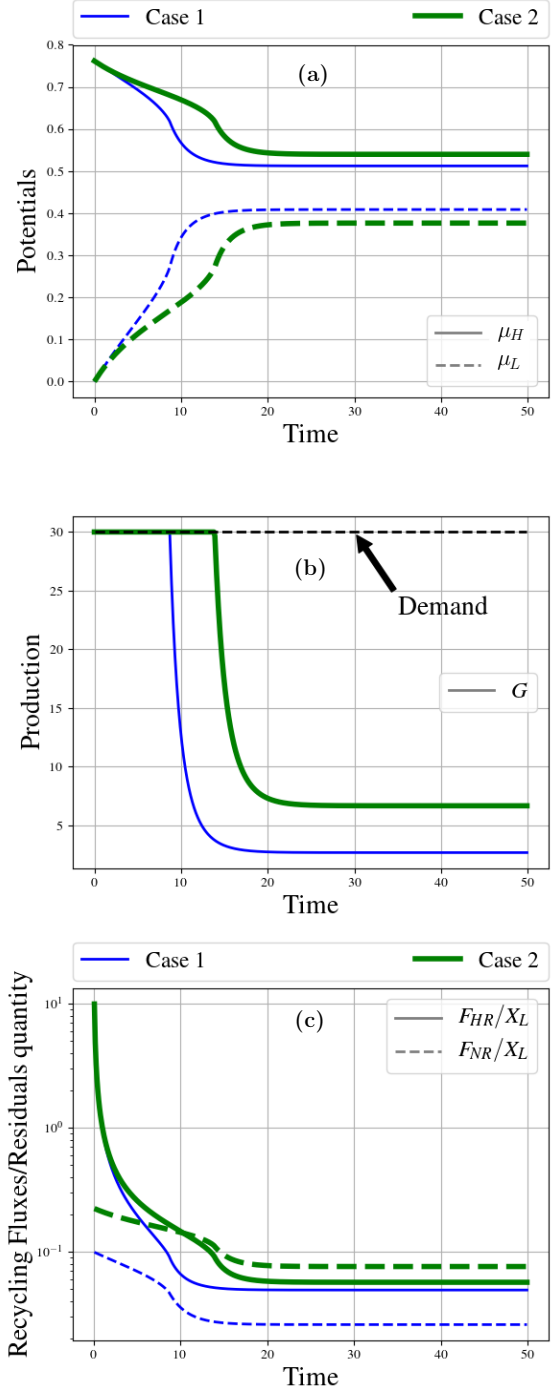

	\begin{center}
		\begin{tabular}{c}
			\includegraphics[width=1\linewidth]{figures/figure6a.pdf}	\\
			\includegraphics[width=1\linewidth]{figures/figure6b.pdf}	\\
			\includegraphics[width=1\linewidth]{figures/figure6c.pdf}
		\end{tabular}
		\captionof{figure}{Resource case studies: Friction impact. From left to right. $a$-Time evolution of high and low potentials ($\mu_H$ and $\mu_L$) for weak (Case 1) and strong (Case 2) friction.
			$b$-Time evolution of production of goods (flux $G$) for weak
			(Case 1) and strong (Case 2) friction. 
			$c$-Time evolution of normalized recycling fluxes ($F_{HR}$ and $F_{NR}$) for weak (Case 1) and strong (Case 2) friction. }
		\label{fig:frictionimpact}
	\end{center}
\end{figure}
Production friction, $R_{P}$, or recycling friction, $R_{R}$, can vary across time. Absent investment, their value increases over time, as a result of capital erosion. Conversely, their value may decrease if investments are made that lead to an increase in capital through the improvement of production facilities. Changing $R_{R}$ does not have a significant impact since we run at $J_{R}^\text{max}$ which adapts to the value of $R_{R}$. We hence choose to consider only the variations of $R_{P}$. We consider two extreme values for $R_{P}$ : very low $R_{P}=4.10^{-5}$ (case 1) and high $R_{P}=0.1$ (case 2). These values were found to yield consistent results.

A low friction, $R_{P}$,  (blue curves of Figs.\,\ref{fig:frictionimpact}-$c$) means a good production system, which runs effectively. It is thus capable of attaining the required demand by producing enough. However, this satisfactory production intensity is accompanied by corresponding consumption and waste disposal. As can be seen in Figure\,\ref{fig:frictionimpact}-$b$. However, after a short phase of strong production in adequation with demand, at time 13, the potentials are pinched, the waste reservoir saturates, the reservoir of initial resources empties and production suddenly collapses. A minute production still remains, thanks to recycling which ensures a slight renewal of resources (Fig.\,\ref{fig:frictionimpact}-$c$). This is hampered by the large amount of waste present: the low-potential reservoir level becomes high. Conversely, a high friction $R_P$ (green curves of Figs\,\ref{fig:frictionimpact}-$a$ to\,\ref{fig:frictionimpact}-$c$) corresponds to a system in poor condition. Production is difficult (see Fig.\,\ref{fig:frictionimpact}-$b$) and the potentials only weakly pinch  (Fig.\,\ref{fig:frictionimpact}-$a$). 
Consequently, consumption of raw resources and  production of waste are both low. This allows a good recycling process (Fig.\,\ref{fig:frictionimpact}-$c$), since the
waste reservoir does not fill (saturation is not reached). A system in poor condition is equivalent to a slowly operating system, which allows time for recycling. 
As a result, production is reduced, but less, so than for low friction. We guess from the very different physics of these two extremes that an optimal value for $R_P$ should hence be located between the two extreme values. There is therefore an optimum that can be considered as a form of impedance matching.

\section{ The anthroposhere } \label{SFC}

We next deploy the internal economic dynamics to the thermal engine considered thus far from a termodynamical viewpoint.

\subsection{Stock-flow consistency}

Let us consider a simple SFC set-up with only two sectors: households and firms, hence excluding banks, the government sector and the rest of the world. Say's law is also assumed in the sense that economic supply of final goods/services is postulated to be always absorbed by aggregate demand.\footnote{For a more general version of the economic model with households' debt, see \cite{giraud_household_2021}
, without Say's law, and with an endogenous $\nu$ and money, see \cite{Giraud-Dossetto} 
}

\renewcommand{\arraystretch}{1.1}
\setlength{\tabcolsep}{0.5cm}
\setlength{\intextsep}{12pt plus 2pt minus 10pt}
\begin{figure*}
	\begin{center}
		\begin{tabular}{cccccc}
			\hline 
			
				& Households 
					&  \multicolumn{2}{c}{ Firms}   
						& Sum 
							&\\ 
			\hline 
			
			\textbf{Balance sheet}   
				& 
					& 
						&  
							&  
								&\\ 
			
			Capital 
				&   
					& \multicolumn{2}{c}{$p\,K$ }   
						& $p\,K$ 
							&\\ 
			
			\hline 
			\hline 
			
			\textbf{Transactions} 
				&  
					& current 
						& capital 
							&  
								& \\ 
			Consumption 
				& $-p\,C$ 
					&  $p\,C$ 
						&  
							&  
								& \\ 
			Investment 
				&  
					& $p\,I$ 
						& $-p\,I$  
							& 
								& \\ 
			Wages 
				& $w\,L$ 
					&  $-w\,L$ 
						&   
							&  
								&\\
			
			\hline 
			Financial Balances 
				& $S^h$ 
					&  $\Pi$ 
						&  $-p \, I $ 
							& 
								& \\ 
			\hline 
			
		\end{tabular} 
		\captionof{table}{Balance sheet, transactions in the economy}
		\label{matrix}
	\end{center}
\end{figure*}


\subsection{Goodwin set-up}

In the present section, we embed our elementary sheet approach within an economic framework, namely the Goodwin model, drawing on  \cite{Grasselli,Coping}. The macroeconomic modules are governed by the following system of differential equations: 
\begin{eqnarray}
\bigdot{\omega}_{G} &=& \omega_{G} \; \left[\phi(\lambda_{G}) - \alpha \right] \label{omega}\\
\bigdot{\lambda}_{G} &=& {\color{black}{\lambda_{G} \; \left[\frac{1-\omega_{G}}{\nu} - \alpha - n - \delta \right]}} \label{lambda}
\end{eqnarray}
where $\nu:=K/Y= 2.89$ stands for the (constant) capital-to-output ratio while $\omega_{G}:=W/(p\,Y)$ and $\lambda_{G}:=L/N$ denote respectively the wage share and the employment rate: $W{\color{black}{_G}}$ stands for the wage bill of the economy, $p{\color{black}{_G}}>0$ is the current price level, and $p{\color{black}{_G}}\,Y{\color{black}{_G}}$ refer to the money value of current output\footnote{For simplicity, $Y$ can be thought of as real {\sc gdp} or, else, National income deflated by inflation.}, $L{\color{black}{_G}}$, the amount of labour performed by the human population, $N$. Harrod-neutral labor productivity, $a\geq 0$, grows exponentially at the rate $\alpha>0$:\footnote{See \cite{Coping} for details.}
\begin{eqnarray}
\alpha &=& \frac{\bigdot{a}}{a} = 2.26 \; 10^{-2}. 
\end{eqnarray}
The growth rate of labour forces, $N$, is $n$:
\begin{eqnarray}
n &=& \frac{\bigdot{N}}{N} = q \; \left( 1-\frac{N}{P^N}\right) = 9.7 \; 10^{-3},\label{popN}
\end{eqnarray}
\noindent with $P^N = 7.059 \; 10^9$, the upper limit of the population, and $q = 2.7 \; 10^{-2}$, its speed of growth \cite{Coping}. The parameter, $\delta = 6.25\; 10^{-2}$, is the depreciation rate of capital, which obeys the standard rule of accumulation expressed in real terms:
\begin{eqnarray}
\bigdot{K}{\color{black}{_G}} &=& I_G - \delta \, K{\color{black}{_G}} \label{Kapital}
\end{eqnarray}
Eventually, the function $\phi(\cdot)$ refers to the short-run Phillips curve. It drives the growth rate of real wages as a function of the  employment rate:
\begin{eqnarray}
\bigdot{w}_{G} &=& w_{G} \; \phi(\lambda_{G}), \label{wage}
\end{eqnarray}
and reads:
\begin{eqnarray}
\phi(\lambda)  &=& \phi_0 + \phi_1 \lambda,
\end{eqnarray}
\noindent whose parameters, $ \phi_0 = -0.73 $ and $\phi_1 = 1.08$, have been empirically estimated on the world economy for the past three decades in \cite{Coping}. 

Following \cite{LavGod} among others, prices are assumed to be given by a fixed margin on direct unitary costs. Direct costs consist only of wages in this simplified model.\footnote{Admittedly, further work will need to incorporate the endogenous prices of E\&M.} The price of the consumption goods and services, $p{\color{black}{_G}}$, is driven by a markup $m>1$, wages $w{\color{black}{_G}}\,L{\color{black}{_G}}=W{\color{black}{_G}}$ and real output, $Y$. It reads:
\begin{eqnarray}
p{\color{black}{_G}} &=&  (1+m) \; \frac{W{\color{black}{_G}} \, L{\color{black}{_G}}}{Y{\color{black}{_G}}} \label{price}
\end{eqnarray}
{\color{black}{The output flow, $Y_G$, depends on the wage share, $\omega_G$, the output capital ratio $\nu$ and capital depreciation $\delta$ see \cite{Grasselli,Coping}. Its evolution is given by:
\begin{eqnarray}
\bigdot{Y}{\color{black}{_G}} &=& Y{\color{black}{_G}} \; \left(\frac{1-\omega_{G}}{\nu} - \delta \right). \label{output}
\end{eqnarray}
}}
The nominal net profits of private firms, $\Pi{\color{black}{_G}}$, depends on the difference between nominal output and the global cost of production (i.e., the wage bill):
\begin{eqnarray}
\Pi{\color{black}{_G}} &=& p{\color{black}{_G}} \, Y{\color{black}{_G}} - w_{G} \, L{\color{black}{_G}}. \label{profitgk}
\end{eqnarray}
Investment, $I{\color{black}{_G}}$, is an increasing function of profits normalized by nominal output: $\Pi{\color{black}{_G}}/(p{\color{black}{_G}} \, Y{\color{black}{_G}})=1 - \omega{\color{black}{_G}}$. For simplicity, and following \cite{Grasselli}, let us assume that it is given by:
\begin{eqnarray}
I{\color{black}{_G}} &=& Y{\color{black}{_G}} \; (1-\omega{\color{black}{_G}}). \label{Invgk}
\end{eqnarray}

Table \ref{initialvalue} provides the initial values of the economic system.\footnote{Wages and the real output level are expressed in 2010 US \$.}

\begin{figure}
	\begin{center}
		\begin{tabular}[width=0.5\textwidth]{|c|c|c|}
			\hline 
			
				& Description 
					& Initial value \\ 
			\hline 
			$\omega$ 
				& Wage share  
					& $0.58$ \\ 
			\hline 
			$\lambda$ 
				& Employment rate  
					& $0.69$ \\ 
			\hline 
			$N$ 
				& Workforce 
					& $4.55 \; 10^9$ \\ 
			\hline 
			$w$ 
				& Wage 
					& $11.98$ \\ 
			\hline  
			$Y$ 
				& Real output level 
					& $64.45 \; 10^9$ \\ 
			\hline 
		\end{tabular} 
		\captionof{table}{Initial values of macroeconomic simulations}
		\label{initialvalue}
	\end{center}
\end{figure}

\setlength\abovecaptionskip{0.0ex}
\begin{figure*}
	\begin{center}
		\newcolumntype{M}[1]{>{\centering\arraybackslash}m{#1}}
		\begin{tabular}{|c|m{2cm}|m{2cm}|m{2cm}|m{2cm}|m{2cm}|}
			\hline 
			Scenario 
				& G 
					& 1 
						& 2 
							& 3 
								& 4 \\ 
			\hline 
			$X_T$ 
				& - 
					& $\mathbf{10^8} $
						& 100.0 	
							& 100.0 
								& 100.0 \\ 
			\hline 
			$R_{P}$ 
				& - 
					& 0.001 
						& 0.001
							& 0.001 
								& \textbf{0.1} \\ 
			\hline Features
				& Goodwin pathway 
					& Infinite resources 
						&  Standard example 
							&\textbf{Without forced recycling}
								& High friction  \\ 
			\hline 
		\end{tabular} 
		\captionof{table}{Parameters of the macroeconomic scenario}
		\label{Scenario}
	\end{center}
\end{figure*}

\subsection{Connecting the bio- and the anthroposphere}

\begin{figure*}
	\centering
	\includegraphics[width=1\textwidth]{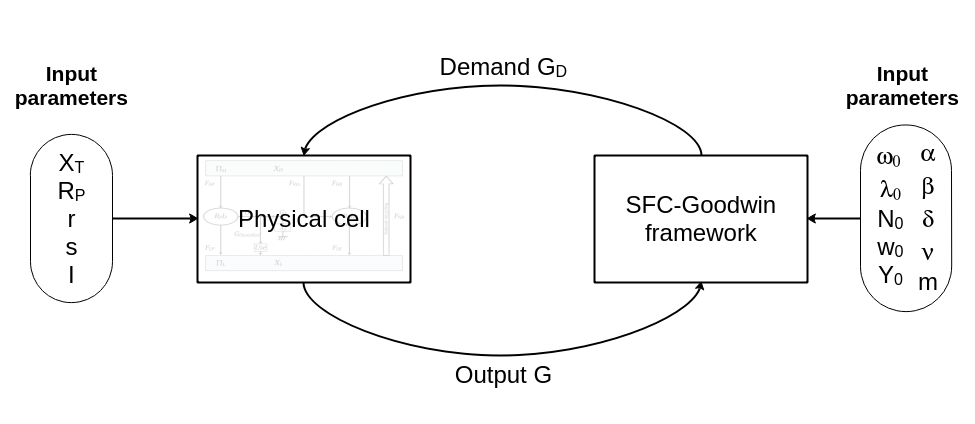}
	\captionof{figure}{Physical sheet into a SFC-Goodwin framework}
	\label{scenariolienG} 
\end{figure*}

The global schematic of the interplay between the physical sheet described in section \ref{conversion} and the economic framework of section \ref{SFC} is represented in Fig.\,\ref{scenariolienG}. Starting from $Y(t)$ at time $t\geq 0$, the aggregate demand is $G_D(t):=Y(t)$. The physical sheet receives this demand from the anthroposphere and delivers an output, $G(t)$, which depends upon the initial quantity of (natural) resources, the friction, and the recycling rate. In the same way as in section\,\ref{casestudy}, $J_P$ is given by \eqref{optimal-intensity}. Eventually, $G_D$ evolves according to  \eqref{output}. The friction is an endogenous function of the capital stock. The larger is capital, the more slowly it degrades, 
\begin{eqnarray}
R_{P} &=&  \frac{ R_{P0} \, K_0 }{ K } + 4.0 \; 10^{-5} 
\label{frictionphy}
\end{eqnarray}

\begin{figure}
	\begin{center}	
		\begin{tabular}{c}
			\includegraphics[width=0.4\textwidth]{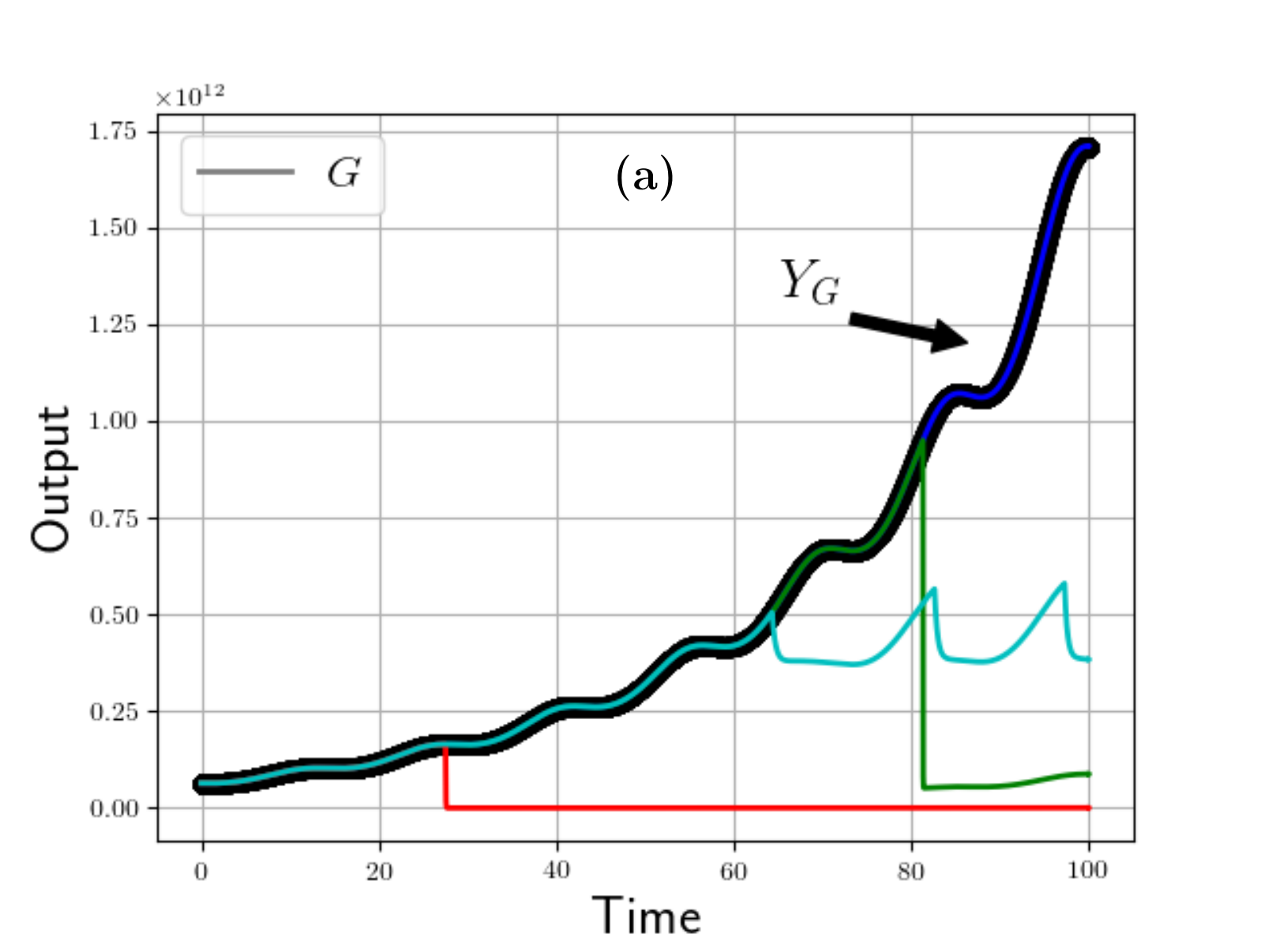} \\
			\includegraphics[width=0.4\textwidth]{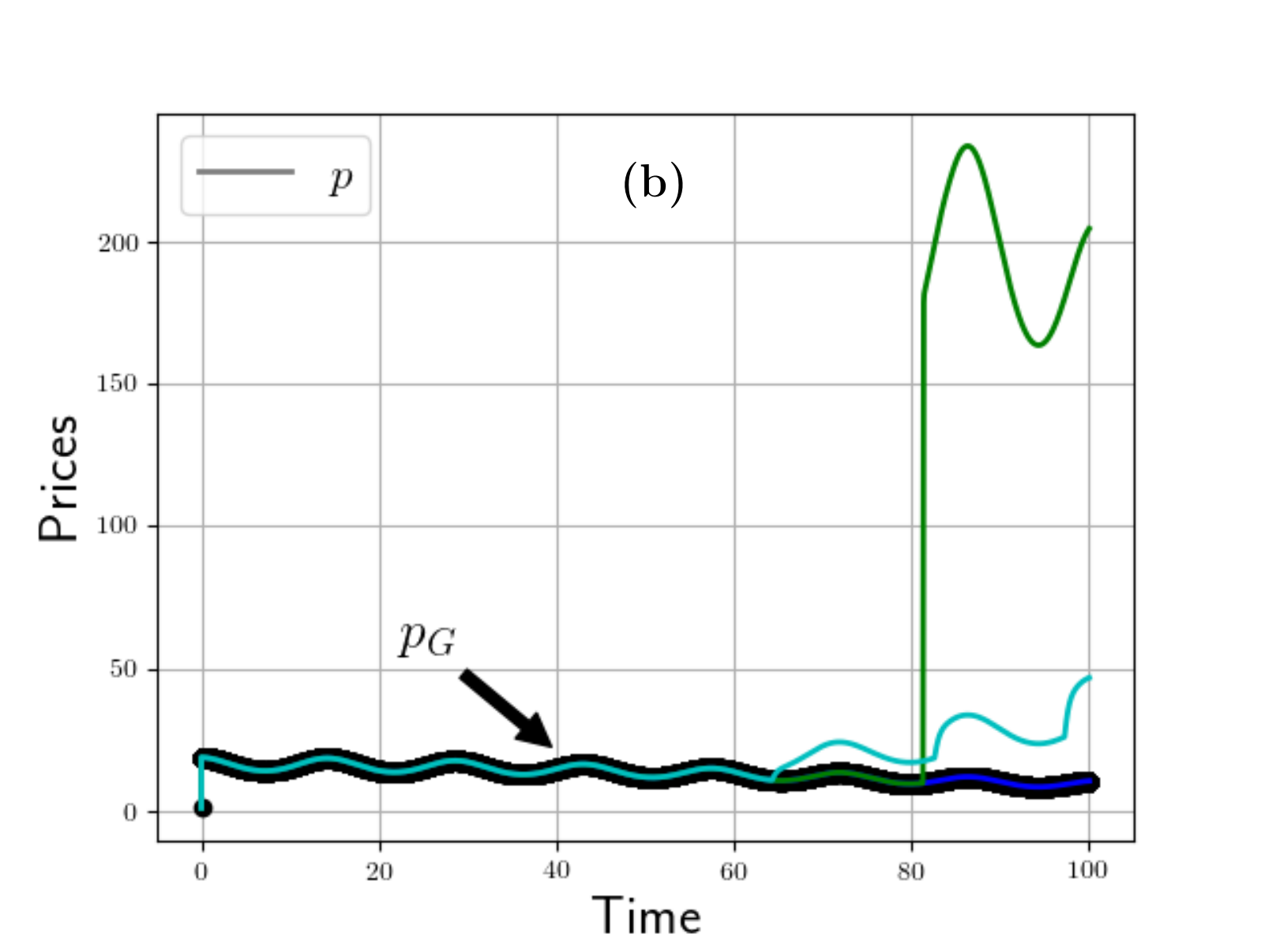}  \\
			\includegraphics[width=0.4\textwidth]{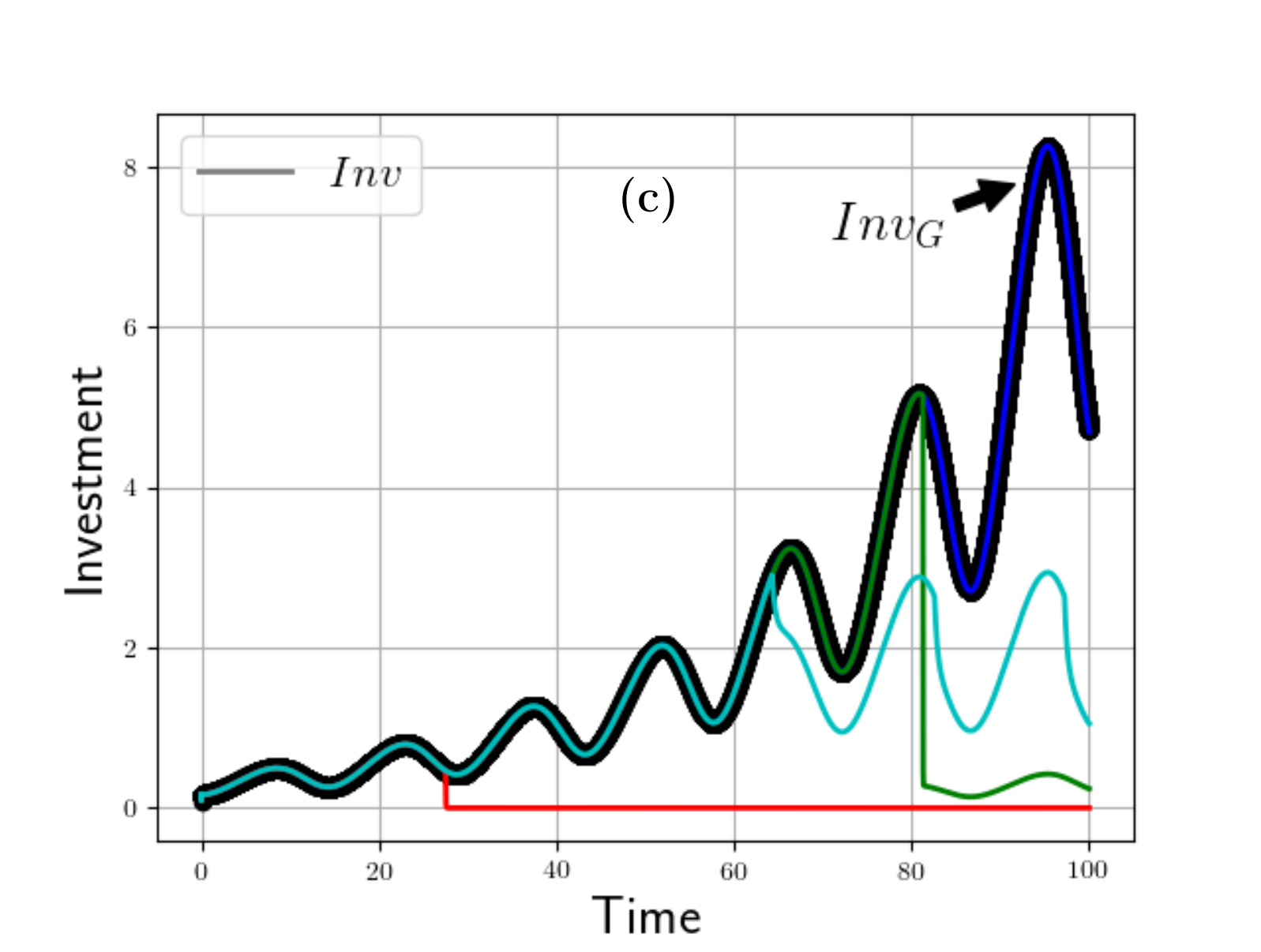}
		\end{tabular}
		\captionof{figure}{Macroeconomic simulations: Economics outputs. From left to right, $a$-Time evolution of production of output, $b$-Time evolution of price and $c$-Time evolution of price. }
		\label{fig:baseline}
	\end{center}
\end{figure}

The initial values presented in table\,\ref{initialvalue} come from \cite{Coping}. The purpose of our simulation is to analyse the effects of taking into account the amount of resource, the recycling and the friction in a SFC framework of the Goodwin type. The baseline scenario represents a Goodwin pathway. This provides a benchmark, absent resource, recycling, friction consideration. In particular, in our baseline scenario output grows over time such as an exponential oscillating curve. This is due to the form of the equation. All economic quantities increase over time. Results of the baseline scenario are shown in Fig.\,\ref{fig:baseline}-$a$,\,\ref{fig:baseline}-$b$,\,\ref{fig:baseline}-$c$ and are represented by the black line. Since resources are considered infinite, economy expands without limit. A crucial question is how the baseline results are modified when we add our physical sheet and in a second time how parameters impacts the results. In what follows, we present a sensitivity analysis that enlightens the impact of a finite quantity of resource \g{and the effect of recycling}. Based on section\,\ref{casestudy} we have selected to modify the following parameters : $X_T$, $R_P$, \g{with or without recycling}.      

\begin{figure*}
	\begin{center}
		\begin{tabular}{cc}
			\multicolumn{2}{c}{\includegraphics[width=.8\textwidth]{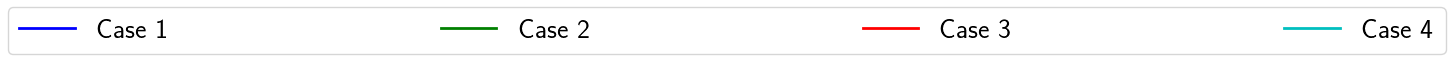}} \\
			\includegraphics[width=.4\textwidth]{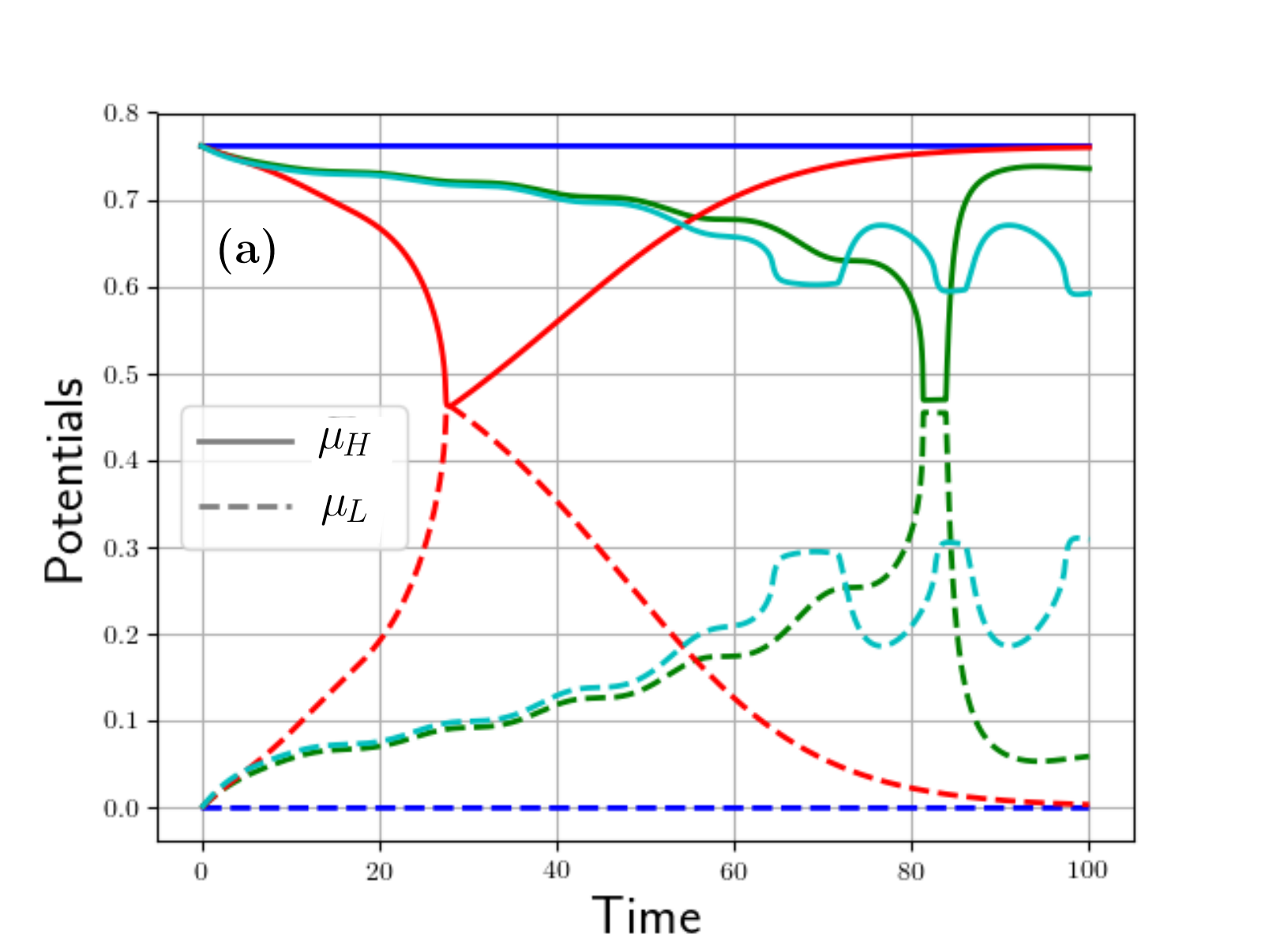} &
			\includegraphics[width=.4\textwidth]{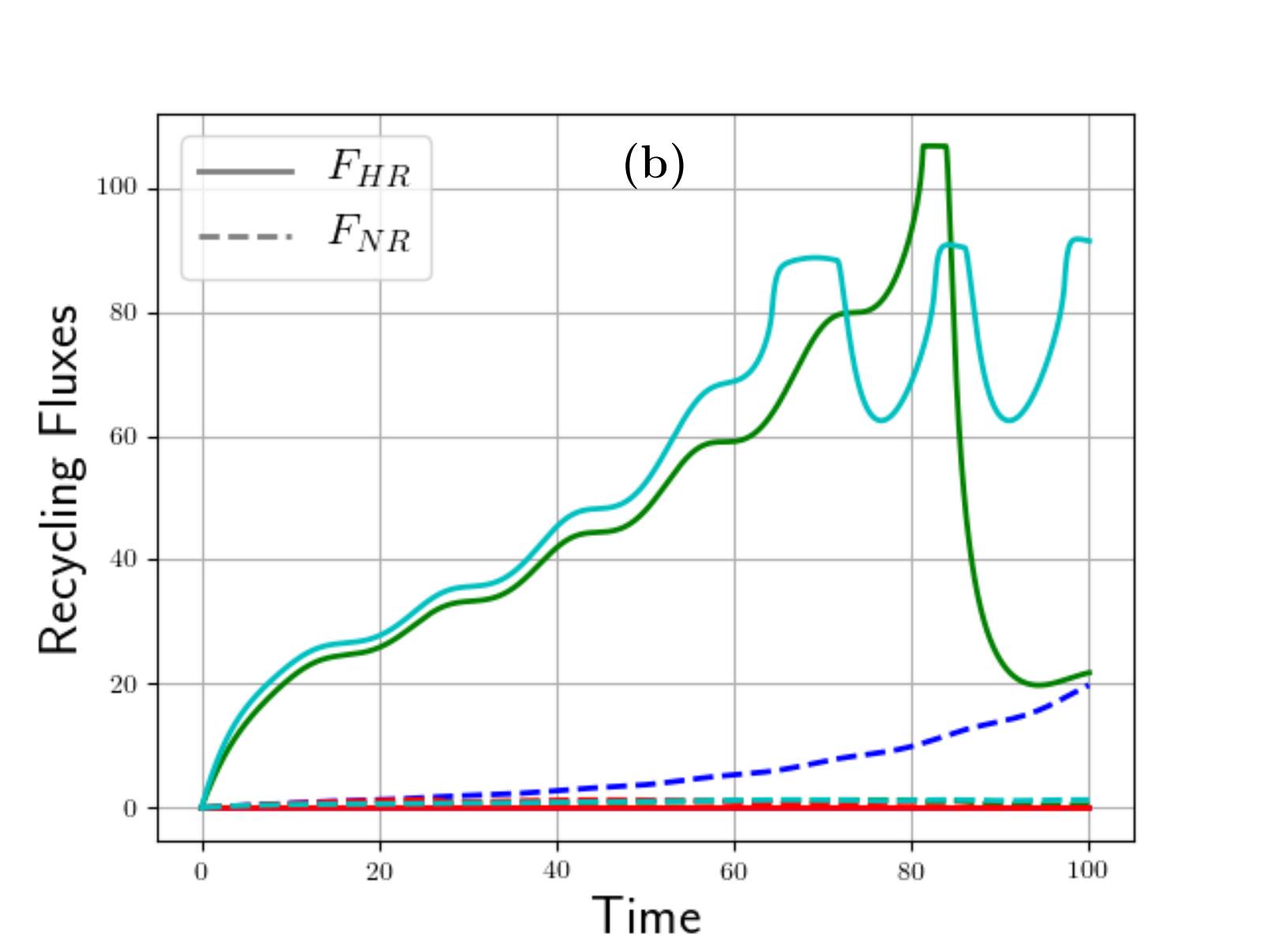} \\
			\includegraphics[width=.4\textwidth]{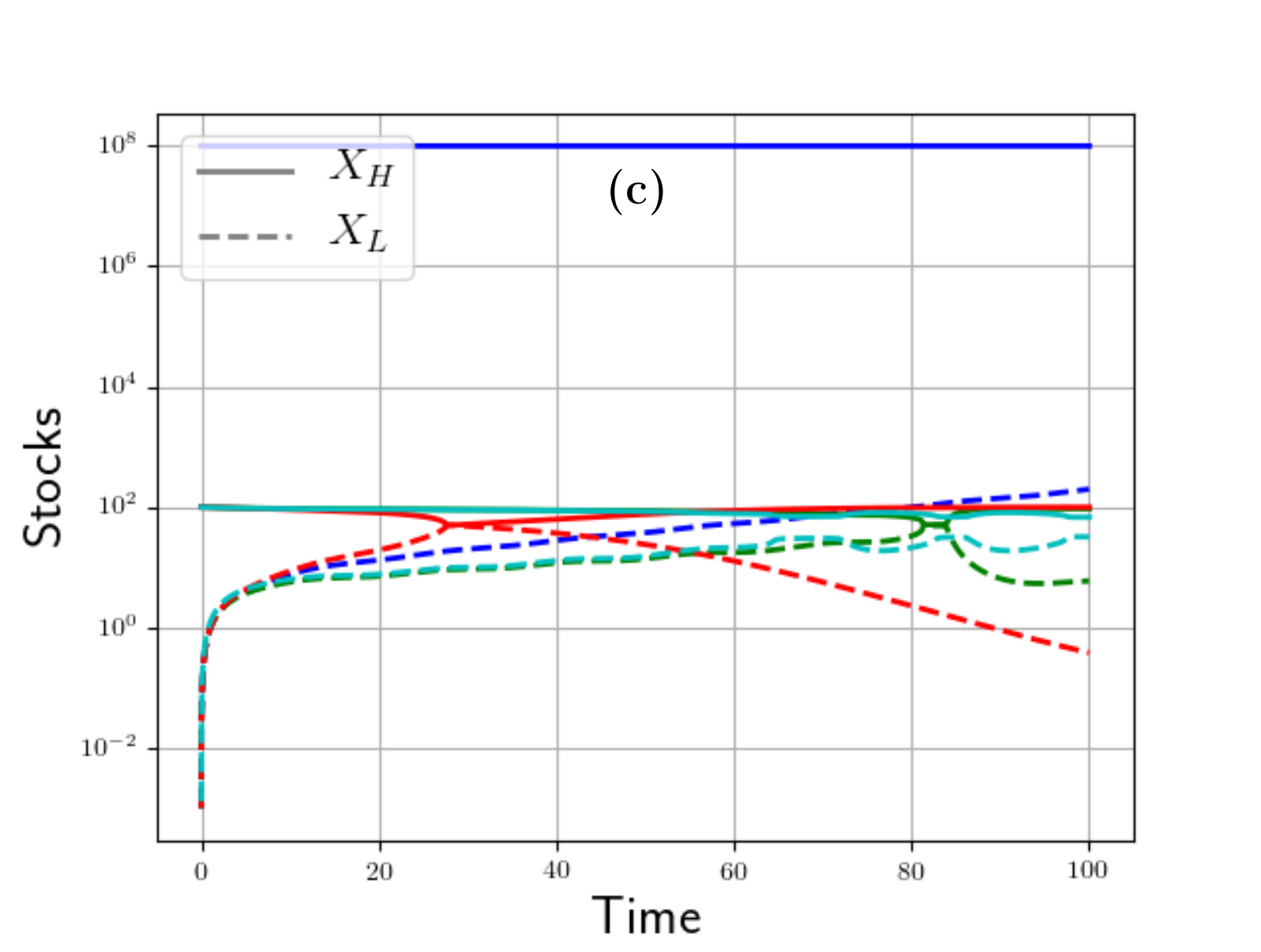} &
			\multirow[c]{2}{*}[4.5cm]{\includegraphics[width=.43\textwidth]{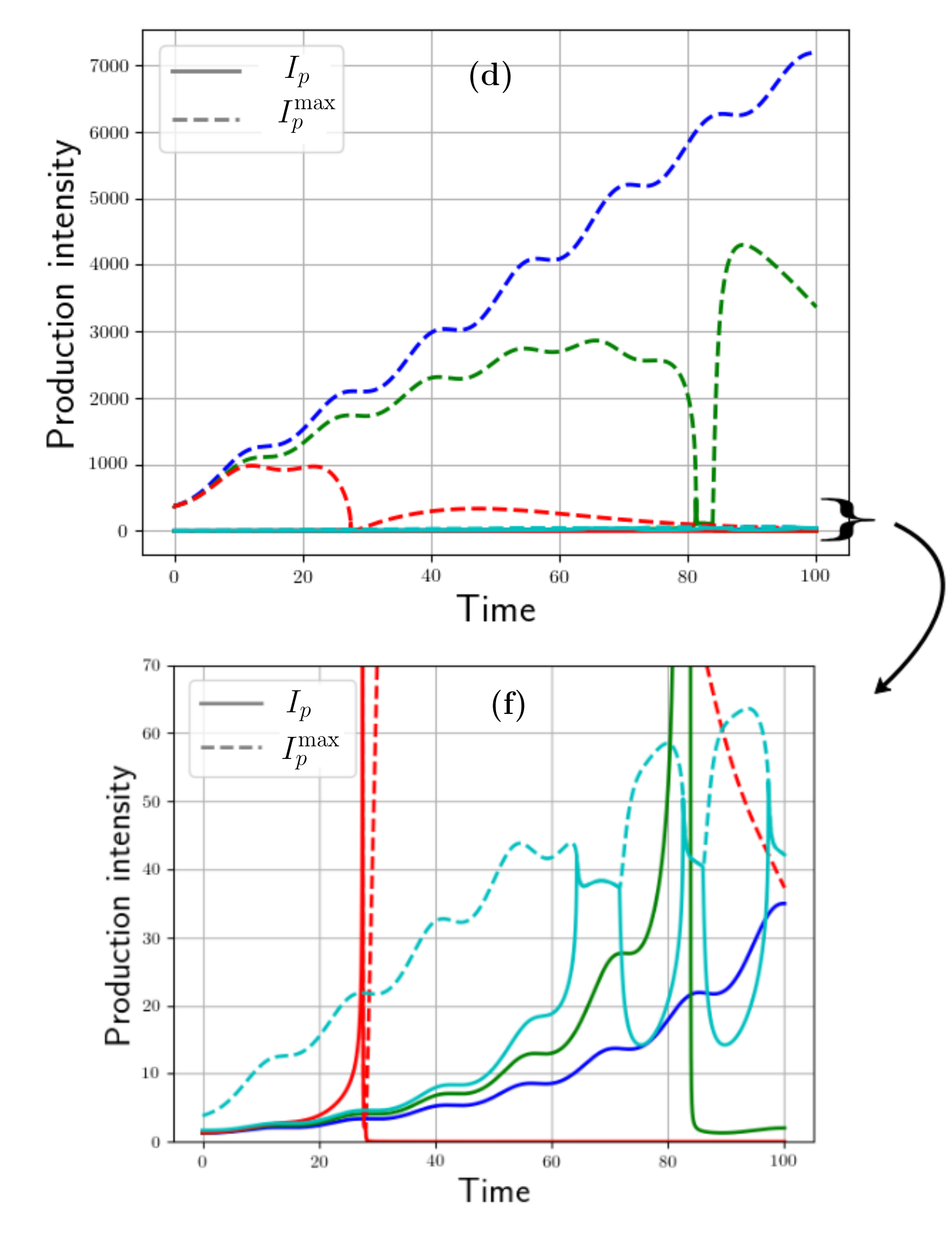}}\\
			\includegraphics[width=.4\textwidth]{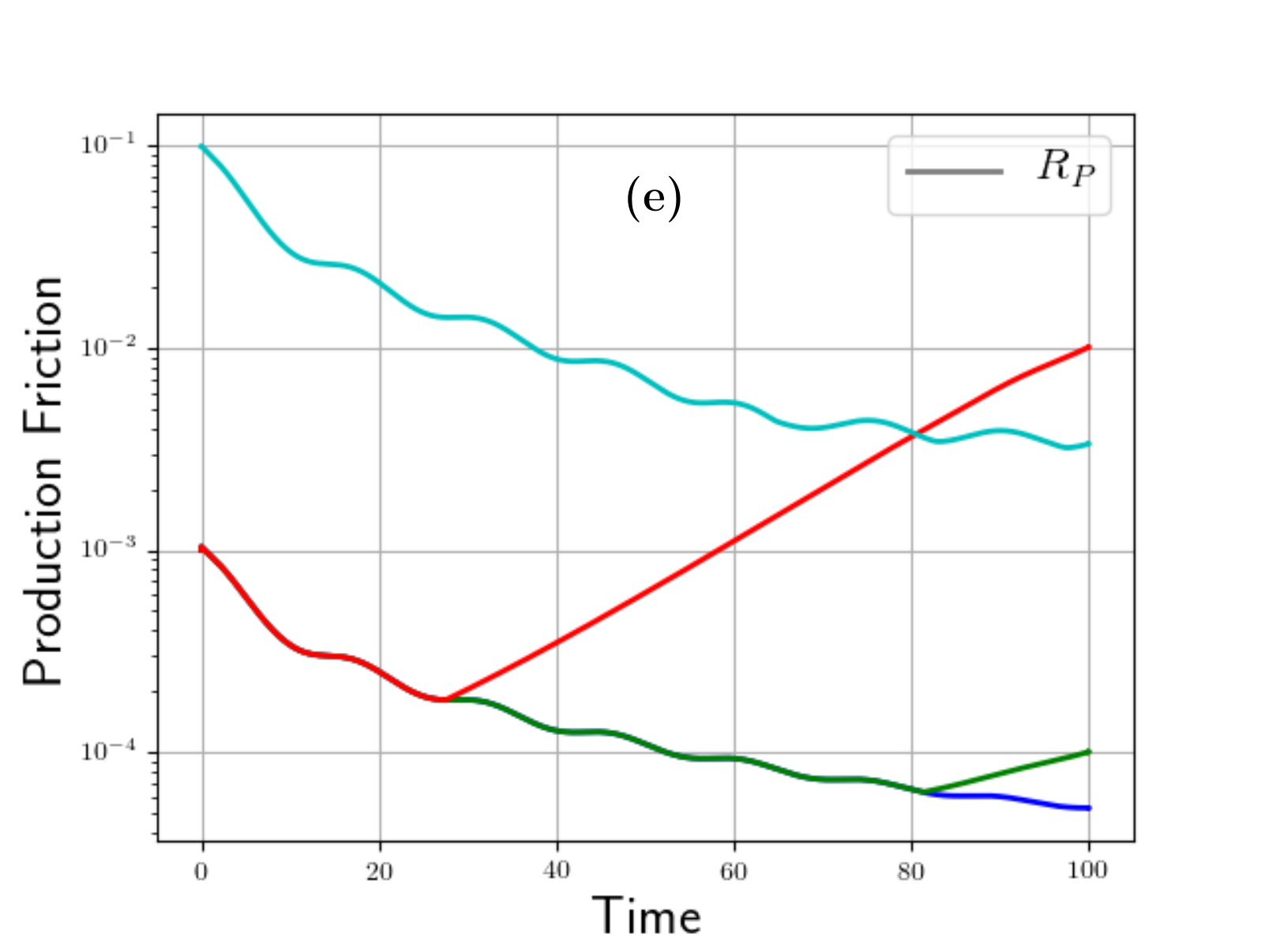} &
		\end{tabular}
	\end{center}
	\caption{Macroeconomic simulations: Resource outputs. From left to right, $a$-Time evolution of high and low potentials ($\mu_H$ and $\mu_L$). $b$-Time evolution recycling fluxes ($F_{HR}$ and $F_{NR}$). $c$-Time evolution of stocks. $d$-and $f$-Time evolution of production intensity. $e$-Time evolution of production friction. } 
	\label{fig:5scenarios}
\end{figure*}

We consider 5 classes of scenarios. We take the mark--up value $m =0.2$. Figures\,\ref{fig:5scenarios} and \ref{fig:baseline} present the deterministic trajectories of our model for the five scenarios under consideration. Table\,\ref{Scenario} wraps up the varying parameters of our four scenarios, the scenario $G$ being only a pure Goodwin framework (and hence it does not need any parameters). The first scenario (blue curve) considers a vast quantity of primary resource with the idea to have a quantity of resource as infinite. The second scenario (green cruve) is a simple example of considering a relatively ``small'' quantity of $X_T$ as a finite world. The third (red curve) and the fourth (cyan curve) are variations of the second. The third scenario (red curve) runs without any forced recycling ($n_R = 0$). The fourth scenario runs with an extreme upper production friction coefficient $R_P$. As in paragraph\,\ref{casestudy}, we run at $J_{R}^\text{max}$ and $J$ is calculated to obtain the requested output. The demand equation, Eq.\,\eqref{output}, follows the black curve until the demand is no longer satisfied. Then, the demand depends on the goods actually produced. For scenario 1 (blue curve), potentials remain constant over time and their difference is maximum. This is due to the fact that since the quantity of resource is very large, the consumption is low, compared to the total quantity.  As a consequence, production is able to follow the production given  by the pure Goodwin model. Natural recycling is high because of the high $X_H$ quantity. On the other hand, a large quantity of resources means a large reservoir, therefore a low potential, low even if waste is produced. The result is an absence of forced recycling  (see Fig.\,\ref{fig:5scenarios}-$b$). Investment and price (Figs.\,\ref{fig:baseline}-$c$ and\,\ref{fig:baseline}-$b$) are the same than in the Goodwin case. Finally, $R_P$ decreases due to the
increase of capital (Fig.\,\ref{fig:5scenarios}-$e$). The Goodwin framework actually represents a world with infinite resources. 

Scenario 2 (green curve) represents a world where resources are finite. The scenario is able to closely reproduce Goodwin production evolution for a while,
 but at a certain time, the production suddenly decreases and potentials are strongly pinched. We consumed the stock of initial resources and saturation of forced recycling is reached.  Investment (Fig.\,\ref{fig:baseline}-$c$) follows the production trend. The potentials then begin to evolve again, because $\omega$  is in
  an ascending phase, and that results in a decrease in demand, in comparison of what it would have been at a lower $\omega$. The gradual reduction in demand allows the system to ``recover''  --- especially thanks to natural and forced recycling, which regenerates very well the resource, because consumption is low --- and eventually no longer operates at maximum intensity (see Fig.\,\ref{fig:5scenarios}-$d$).  As a result of the drop in production, prices rise sharply (see Fig.\,\ref{fig:baseline}-$b$), investment
decreases (see Fig.\,\ref{fig:baseline}-$c$) and the state of the production system degrades (see Fig.\,\ref{fig:5scenarios}-$e$).

Scenario 3 exhibits a case where there is no forced recycling, the resource is renewed only through natural recycling. The stock of resources is first consumed until exhaustion. Production follows the Goodwin scheme at the beginning then collapses. Potential and stocks follow distinct developments: the potential goes up fairly quickly, while the stock of initial resources regenerates slowly (see Fig.\,\ref{fig:5scenarios}-$c$). The resource level remains fragile. Production remains low (see Fig.\,\ref{fig:baseline}-$a$) and continues only through natural recycling. As production has dropped, investment as well as the state of the system have also deteriorated. 

Scenario 4 has the maximum value of $R_P$ that yields coherent result. At early times, production follows the Goodwin production curve but then, decreases because of the pinch-off of potentials. Similarly to the first case study, difference of potentials is higher at the end than with a lower $R_P$. As consequence, production is higher after the decrease in production of scenario 2. 
The potentials oscillate because $J_P^\text{max}$ is sometimes reached: the value of $J_P^\text{max}$ depends heavily on the potentials and of the friction, which itself   depends on the oscillating capital.  When capital oscillates,  friction $R_P$ then $J_P^\text{max}$ and flows, stocks and potentials oscillate in return. 
Prices rise gradually as a result of the successive increase and decrease of the level of  produced goods.

\section{Conclusion}
The aim of the paper is to propose a coherent formalism of the interactions between the thermodynamics of the physical world and the macrodynamics of the economic sphere. Three main features are integrated into our model: $(i)-$the quantity of resources (energy and matter) is preserved throughout and  reservoirs of initial resources and wastes are made explicit; $(ii)-$the resources being available in finite quantities, only recycling allows a perennial economy; $(iii)-$the thermodynamically inexorable degradation of energy and matter being made explicit, recycling requires upgrading the quality of wastes, which, in turn, costs energy. 

The strategy consists of building a unitary sheet, based on a thermodynamic conversion machine scheme. Our three-sphere model --- one production sphere and two recycling (natural and forced) zones ---, enables to understand how the economic metabolism uses resources to produce goods/services and to recycle the waste products. Production characteristics vary according to the metabolisation intensity. The intensity corresponding to  maximum possible production rapidly depletes the resource, so that the production regime is physically constraint to quickly evolve from high to very low intensity, while the amount of waste accumulates. Conversely, a moderate intensity leaves time to the resource to renew itself and ultimately produces more useful work than an initially strong operating intensity. Hence, an economic regime emerges, which allows a moderate but sustainable dissipation of energy and degradation of matter. The choice of the optimal operating intensity of such a thermodynamical system in general goes far beyond the simple search for maximum efficiency or maximum production. In fact the variety of operating points of this system is very large. However, three main points can be identified. The first is the maximum yield point, which ensures that the production to resource ratio is maximized. It should be noted that this point in no way means that the resource is used sparingly. The second point is the point of maximum production, with no further concern for resource conservation or waste production. The third and last point concerns minimization of the production of waste.  It never coincides with the two previous ones, \cite{herbert_thermodynamics_2020}. To this end, a distinction must be made between systems that can be described as animate (including all living bodies), as opposed to inanimate systems \cite{goupil2020}. The former exhibit a minimum operating point, called the basal point, which corresponds to the point of rest, below which metabolic functioning cannot be maintained. In other words, unlike inanimate ones, resource consumption by animate systems can never be zero. This results in the emergence of the particular intensity point, at which the production of waste during a given process is minimized. This point never coincides either with the point of maximum power or with the intensity level which maximizes efficiency.
\footnote{This point can be understood, for example, in the choice of speed of movement of an animal over a given distance. At high speeds the dissipation is important, but at very low speeds the operating time is greater, and the same applies to the basal consumption in the background. We can therefore see that there is an intermediate speed which is more favorable, see \cite{herbert_thermodynamics_2020}.} 

The second important parameter is the friction which characterizes the state of the economic metabolism. For two systems considered at their maximum production, a system in ``good shape'' will produce efficiently, \textit{i.e.} with a good ratio of  production of useful work over energy dissipation and matter degradation. However, its major consequence will be to consume more resources, to 
produce more useful work and to exude more wastes. 
A system in worse condition, will consume a lower flow of E\&M, since high friction prevents of the the potential gradient from reducing, 
will reject less wastes and leave time for the resource to renew itself. 

Next, we connect our physically-oriented sheet with an economic set-up of the SFC type, fleshed out with a Goodwin-like dynamics. By contrast with a vast majority of the economic literature (including the seminal Goodwin dynamics), we do not postulate that resources are infinite, so that it is impossible to produce indefinitely with no recycling. We show that perennial paths consist in a moderate use of resources and intensive recycling. This contrasts both with the 
``growth imperative'' which argues that {\sc gdp} growth should be maximized (on the alleged argument that more growth will allow solving the problems created by growth itself) but also with a degrowth perspective.  The  sheet structure makes it possible to complexify the picture, by introducing as many sheets as required. On the other hand, our thermodynamical modelling of resources is independent of the macroeconomic model used and can be easily coupled with other alternative models. 
 
This paper paves the road to a number of further areas of research. Let us mention a few of them. First, an immediate question arises: is it possible, within the setting of this paper, to identify the basal point of the (obviously living) anthroposphere? Second, throughout the paper, we treated the metabolization intensity as an exogenously given parameter: how should we turn it into an endogenous variable, driven by the internal forces of the economic metabolism (within the constraints imposed by the biosphere)? Third, we chose the admittedly simplistic Goodwin dynamics. How would the whole system work were we to replace this preliminary dynamics by a more sophisticated one? Fourth, for the sake of simplicity, we assumed that the friction is a decreasing function of the stock of capital. This idea, however, deserves more investigation. After all, ``capital'' could be understood as the whole set of procedures, innovations, webs, infrastructures... that enable the economic metabolism to reduce its friction parameter.

\section*{Declaration of Competing Interest}
The authors declare that they have no known competing financial interests or personal relationships that could have appeared to influence the work reported in this paper. 

\section*{Acknowledgements}
This work was funded by Chaire Énergie et prosperité. \footnote{see \url{http://www.chair-energy-prosperity.org/en/}}
 The authors would like to thank \textit{Henri Benisty} and \textit{Louison Cahen-Fourot} for fruitfull discussions.

\begin{figure}[htbp]
	\centering
		\begin{tabular}{cm{.3\textwidth}}
			\textbf{Variable}   & \textbf{Signification} \\
			$X_H$       & Stock of primary resource \\
			$X_L$       & Stock of waste \\
			$X_T$       & Total stock \\
			$X_S$       & Buffer stock \\
			$\mu_H$     & Potential of primary resource \\
			$\mu_L$     & Potential of waste \\
			$F_{HP}$    & Flow of primary resource \\
			$F_{LP}$    & Flow of waste \\
			$G$         & Final goods \\
			$G_S$       & Buffer flow \\
			$G_D$       & Demand of goods \\
			$G_{D}^\text{satisfied}$     & (or $G_D^S$) Satisfied demand \\
			$F_{HR}$    & Flow of recycled resource \\
			$F_{LR}$    & Flow of waste used in recycling\\
			$F_{RIn}$   & Flow of resource used to recycle \\
			$F_{NR}$    & Flow of natural recycling \\
			$R_P$       & Production viscosity \\
			$R_R$       & Recycling viscosity \\
			$J$         & Global Intensity \\
			$J_{P}$     & Production intensity\\
			$J_{P}^\text{max}$  & Maximal intensity of production \\
			$\tau$      & Time constant of intensity of production\\
			$J_{R}$     & Recycling intensity\\
			$J_{R}^\text{max}$  & Maximal intensity of recycling \\
			$J_{P}$     & Production intensity\\
			$n_{P}$ & Production intensity modulation coefficient  \\
			$n_{R}$ & Recycling intensity modulation coefficient \\
		\end{tabular}
	\captionof{table}{Variables}
	\label{varirecap}
\end{figure}


\bibliography{Bibliographie}


\end{document}